\documentclass[aps,pra,twocolumn,export]{revtex4-2} 
\usepackage[utf8]{inputenc}

\usepackage{graphicx}
\usepackage{fullpage}
\usepackage{tikz,tikz-cd} 
\usepackage{amsthm,amsmath,amssymb,amsfonts,mathtools,tensor,mhchem}

\usepackage{bbold} 

\newcommand{\id}{\mathbb{1}}
\newcommand{\myinv}[1]{#1^{\scalebox{0.9}[1.0]{-}1}}

\newcommand{\tr}{{\rm Tr}}
\newcommand{\ket}[1]{| #1 \rangle}
\newcommand{\bra}[1]{\langle #1 |}

\renewcommand{\H}{\mathcal{H}}
\renewcommand{\L}{\mathcal{L}}
\renewcommand{\P}{\mathcal{P}}
\newcommand{\G}{\mathcal{G}}

\usetikzlibrary{decorations.pathreplacing,decorations.markings,calc,3d,positioning,quotes}

\tikzset{
  tensor/.style={
    inner sep = 0.055cm,
    shape = circle,
    draw,
    fill
  },
  t/.style={
    inner sep = 0.03cm,
    shape = circle,
    draw,
    fill
  },
}

\tikzset{
  tensor/.style={
    inner sep = 0.055cm,
    shape = circle,
    draw,
    fill
  }
}

\tikzset{
  hopf2/.style={
    line cap=round,
    line width=1.5mm,
  },
  epsilon2/.style={
   draw = red,
   thin,
   densely dotted,
   rounded corners,
   fill opacity=0.2,
   fill=red,
  },
  every picture/.style = {
    baseline={([yshift=-.5ex]current bounding box.center)},
    font=\scriptsize
  },
  irrep/.style={
    anchor=south,
    font = \tiny,
    inner sep=2pt
  }
}

\tikzset{
  every label/.style = {
    text depth=0pt,
    text height=1ex,
  },
}

\tikzset{
     pics/V1/.style args={#1/#2/#3}{
    code = {
	    \coordinate (-mid) at (-0.4,0);
	    \coordinate (-up) at (0.5,0.3);
	    \coordinate (-down) at (0.5,-0.3);
	    \coordinate (-top) at (0,0.45);
	    \coordinate (-bottom) at (0,-0.45);
	  	    \draw[red] (0,0.3)--(-up);
	    \draw[red] (0,-0.3)--(-down);
	    \draw[red] (0,0)--(-mid);
	      \draw[hopf2] (0,-0.35)--(0,0.35);
	    \node[irrep] at (-0.25,0) {$#1$};
	    \node[irrep] at (0.25,0.3) {$#2$};
	    \node[irrep] at (0.25,-0.3) {$#3$};
    }
  },
       pics/V1u/.style args={#1/#2/#3}{
    code = {
	    \coordinate (-mid) at (-0.4,0);
	    \coordinate (-up) at (0.15,0.3);
	    \coordinate (-down) at (0.5,-0.3);
	    \coordinate (-top) at (0,0.45);
	    \coordinate (-bottom) at (0,-0.45);
	  	    \draw[red] (0,0.3)--(-up);
	    \draw[red] (0,-0.3)--(-down);
	    \draw[red] (0,0)--(-mid);
	      \draw[hopf2] (0,-0.35)--(0,0.35);
	    \node[irrep] at (-0.25,0) {$#1$};
	    \node[irrep] at (0.25,0.3) {$#2$};
	    \node[irrep] at (0.25,-0.3) {$#3$};
    }
  },
      pics/V1p/.style args={#1/#2/#3}{
    code = {
	    \coordinate (-mid) at (-0.1,0);
	    \coordinate (-up) at (0.15,0.3);
	    \coordinate (-down) at (0.125,-0.3);
	    \coordinate (-top) at (0,0.45);
	    \coordinate (-bottom) at (0,-0.45);
	  	    \draw[red] (0,0.3)--(-up);
	    \draw[red] (0,-0.3)--(-down);
	    \draw[red] (0,0)--(-mid);
	      \draw[hopf2] (0,-0.35)--(0,0.35);
	    \node[irrep] at (-0.25,0) {$#1$};
	    \node[irrep] at (0.25,0.3) {$#2$};
	    \node[irrep] at (0.25,-0.3) {$#3$};
    }
  },
       pics/V1m/.style args={#1/#2/#3}{
    code = {
	    \coordinate (-mid) at (-0.1,0);
	    \coordinate (-up) at (0.5,0.3);
	    \coordinate (-down) at (0.5,-0.3);
	    \coordinate (-top) at (0,0.45);
	    \coordinate (-bottom) at (0,-0.45);
	  	    \draw[red] (0,0.3)--(-up);
	    \draw[red] (0,-0.3)--(-down);
	    \draw[red] (0,0)--(-mid);
	      \draw[hopf2] (0,-0.35)--(0,0.35);
	    \node[irrep] at (-0.25,0) {$#1$};
	    \node[irrep] at (0.25,0.3) {$#2$};
	    \node[irrep] at (0.25,-0.3) {$#3$};
    }
  },
    pics/W1/.style args={#1/#2/#3}{
    code = {
	    \coordinate (-mid) at (0.4,0);
	    \coordinate (-up) at (-0.5,0.3);
	    \coordinate (-down) at (-0.5,-0.3);
	    \coordinate (-top) at (0,0.45);
	    \coordinate (-bottom) at (0,-0.45);
	   \draw[red] (0,0.3)--(-up);
	    \draw[red] (0,-0.3)--(-down);
	    \draw[red] (0,0)--(-mid);
	    	    \draw[hopf2] (0,-0.35)--(0,0.35);
	    \node[irrep] at (0.25,0) {$#1$};
	    \node[irrep] at (-0.25,0.3) {$#2$};
	    \node[irrep] at (-0.25,-0.3) {$#3$};
    }
  },
      pics/W1u/.style args={#1/#2/#3}{
    code = {
	    \coordinate (-mid) at (0.4,0);
	    \coordinate (-up) at (-0.15,0.3);
	    \coordinate (-down) at (-0.5,-0.3);
	    \coordinate (-top) at (0,0.45);
	    \coordinate (-bottom) at (0,-0.45);
	   \draw[red] (0,0.3)--(-up);
	    \draw[red] (0,-0.3)--(-down);
	    \draw[red] (0,0)--(-mid);
	    	    \draw[hopf2] (0,-0.35)--(0,0.35);
	    \node[irrep] at (0.25,0) {$#1$};
	    \node[irrep] at (-0.25,0.3) {$#2$};
	    \node[irrep] at (-0.25,-0.3) {$#3$};
    }
  },
      pics/W1D/.style args={#1/#2/#3}{
    code = {
	    \coordinate (-mid) at (0.4,0);
	    \coordinate (-up) at (-0.5,0.3);
	    \coordinate (-down) at (-0.15,-0.3);
	    \coordinate (-top) at (0,0.45);
	    \coordinate (-bottom) at (0,-0.45);
	   \draw[red] (0,0.3)--(-up);
	    \draw[red] (0,-0.3)--(-down);
	    \draw[red] (0,0)--(-mid);
	    	    \draw[hopf2] (0,-0.35)--(0,0.35);
	    \node[irrep] at (0.25,0) {$#1$};
	    \node[irrep] at (-0.25,0.3) {$#2$};
	    \node[irrep] at (-0.25,-0.3) {$#3$};
    }
  },
      pics/W1m/.style args={#1/#2/#3}{
    code = {
	    \coordinate (-mid) at (0.1,0);
	    \coordinate (-up) at (-0.5,0.3);
	    \coordinate (-down) at (-0.5,-0.3);
	    \coordinate (-top) at (0,0.45);
	    \coordinate (-bottom) at (0,-0.45);
	   \draw[red] (0,0.3)--(-up);
	    \draw[red] (0,-0.3)--(-down);
	    \draw[red] (0,0)--(-mid);
	    	    \draw[hopf2] (0,-0.35)--(0,0.35);
	    \node[irrep] at (0.25,0) {$#1$};
	    \node[irrep] at (-0.25,0.3) {$#2$};
	    \node[irrep] at (-0.25,-0.3) {$#3$};
    }
  },
  pics/W1p/.style args={#1/#2/#3}{
    code = {
	    \coordinate (-mid) at (0.1,0);
	    \coordinate (-up) at (-0.1,0.3);
	    \coordinate (-down) at (-0.15,-0.3);
	    \coordinate (-top) at (0,0.45);
	    \coordinate (-bottom) at (0,-0.45);
	   \draw[red] (0,0.3)--(-up);
	    \draw[red] (0,-0.3)--(-down);
	    \draw[red] (0,0)--(-mid);
	    	    \draw[hopf2] (0,-0.35)--(0,0.35);
	    \node[irrep] at (0.25,0) {$#1$};
	    \node[irrep] at (-0.25,0.3) {$#2$};
	    \node[irrep] at (-0.25,-0.3) {$#3$};
    }
  },
    pics/V2/.style args={#1/#2/#3}{
    code = {
	    \coordinate (-mid) at (-0.4,0);
	    \coordinate (-up) at (0.5,0.3);
	    \coordinate (-down) at (0.5,-0.3);
	    \coordinate (-top) at (0,0.45);
	    \coordinate (-bottom) at (0,-0.45);
	    \draw[red] (0,0.3)--(-up);
	    \draw (0,-0.3)--(-down);
	    \draw (0,0)--(-mid);
	    \draw[hopf2,gray] (0,-0.35)--(0,0.35);
	    \node[irrep] at (-0.25,0) {$#1$};
	    \node[irrep] at (0.25,0.3) {$#2$};
	    \node[irrep] at (0.25,-0.3) {$#3$};
    }
  },
  pics/W2/.style args={#1/#2/#3}{
    code = {
	    \coordinate (-mid) at (0.4,0);
	    \coordinate (-up) at (-0.5,0.3);
	    \coordinate (-down) at (-0.5,-0.3);
	    \coordinate (-top) at (0,0.45);
	    \coordinate (-bottom) at (0,-0.45);
	   \draw[red] (0,0.3)--(-up);
	    \draw (0,-0.3)--(-down);
	    \draw (0,0)--(-mid);
	    \draw[hopf2,gray] (0,-0.35)--(0,0.35);
	    \node[irrep] at (0.25,0) {$#1$};
	    \node[irrep] at (-0.25,0.3) {$#2$};
	    \node[irrep] at (-0.25,-0.3) {$#3$};
    }
  },
   pics/W3/.style args={#1/#2/#3}{
    code = {
	    \coordinate (-mid) at (0.55,0);
	    \coordinate (-up) at (-0.5,0.3);
	    \coordinate (-down) at (-0.5,-0.3);
	    \coordinate (-top) at (0,0.45);
	    \coordinate (-bottom) at (0,-0.45);
	    \draw (0,-0.3)--(-down);
	    \draw (0,0)--(-mid);
	    \draw[hopf2,gray] (0,-0.35)--(0,0.35);
	    \node[irrep] at (0.35,0) {$#1$};
	    \node[irrep] at (-0.25,0.3) {$#2$};
	    \node[irrep] at (-0.25,-0.3) {$#3$};
    }
  },
  pics/V3/.style args={#1/#2/#3}{
    code = {
	    \coordinate (-mid) at (-0.55,0);
	    \coordinate (-up) at (0.5,0.3);
	    \coordinate (-down) at (0.5,-0.3);
	    \coordinate (-top) at (0,0.45);
	    \coordinate (-bottom) at (0,-0.45);
	    \draw (0,-0.3)--(-down);
	    \draw (0,0)--(-mid);
	    \draw[hopf2,gray] (0,-0.35)--(0,0.35);
	    \node[irrep] at (-0.35,0) {$#1$};
	    \node[irrep] at (0.25,0.3) {$#2$};
	    \node[irrep] at (0.25,-0.3) {$#3$};
    }
  },
}

\begin{document}

\title{Gauging quantum states with non-anomalous matrix product operator symmetries}

\author{ Jos\'e Garre-Rubio}
\affiliation{\mbox{University of Vienna, Faculty of Mathematics, Oskar-Morgenstern-Platz 1, 1090 Vienna, Austria}}
\author{ Ilya Kull}
\affiliation{University of Vienna, Faculty of Physics \& Vienna Doctoral School in Physics,  Boltzmanngasse 5, 1090 Vienna, Austria}

\begin{abstract}
Gauging a  global symmetry of a system amounts to introducing new degrees of freedom whose transformation rule makes  the overall system observe a local symmetry. 
In quantum systems there can be obstructions to gauging a global symmetry. When this happens the symmetry is dubbed anomalous. Such obstructions are related to the fact that the global symmetry cannot be written as a tensor product of local operators. In this manuscript we study non-local symmetries that have an additional structure: they take the form of a matrix product operator (MPO). We exploit the tensor network structure of the MPOs to construct local operators from them satisfying the same group relations, that is, we are able to localize even anomalous MPOs. For non-anomalous MPOs, we use these local operators to explicitly gauge the MPO symmetry of a one-dimensional quantum state obtaining non-trivial gauged states. We show that our gauging procedure satisfies all the desired properties as the standard on-site case does. We also show how this procedure is naturally represented in matrix product states protected by MPO symmetries. In the case of anomalous MPOs, we shed light on the obstructions to gauging these symmetries.

\end{abstract}

\maketitle

\section{Introduction}

The gauging of a global symmetry, that is, promoting it from global to local by adding new degrees of freedom, the gauge fields, plays a key role in physics: it is the origin of the fundamental interactions in the Standard Model. The gauging of a symmetry is usually prescribed at the level of Lagrangians or Hamiltonians, but gauging global symmetries at the level of states has been previously proposed \cite{Haegeman14}. This is particularly relevant for constructing families of variational states that exhibit the gauged symmetry and can be used to study, for example, lattice gauge theories \cite{Silvi14,Tagliacozzo14,Erez16,Kull17,Kasper20}.

Interestingly, in quantum systems there are global symmetries that cannot be gauged. This situation happens when the symmetry in a theory has a \mbox{'t Hooft} {\it anomaly} \cite{tHooft79}. In condensed matter physics, a natural situation where anomalies arise is at the boundary theories of systems exhibiting symmetry protected topological (SPT) order of a finite group $G$. If the SPT order is non-trivial, the global symmetry is translated at the boundary into an operator that cannot be written as the tensor product of local operators acting on every site (from now on referred to as {\it on-site}). This non-local operator captures the quantum phase of the $d$-dimensional SPT bulk, encoded in an element of  $\mathcal{H}^{d+1}(G,U(1))$ (the $d+1$th cohomology group of $G$), and it corresponds to an anomalous symmetry of the boundary \cite{WenAno13,Chen13,Kapustin14,Else14}. Importantly only non-anomalous symmetries can have unique gapped ground states \cite{Chen13, Thorngren19}.

That picture is very well understood in terms of symmetries in tensor network states \cite{reviewPEPS}. This class of states is defined on lattices by placing local tensors in every site, where the contracted edge indices are the so-called virtual degrees of freedom. Tensor network states have arisen as the main analytical tool to classify SPT phases in one dimension (1D) \cite{Chen11, Schuch11} and two dimensions (2D) \cite{Chen11A,Andras18B} outside zero correlation length points. 
Instrumental to this analysis is that symmetries in tensor network states are encoded at the level of the local tensors and this encoding determines the mapping between the bulk symmetry operators and the boundary symmetry operators acting on the virtual indices \cite{Sanz09,Andras18B,Molnar18A}. In 2D, projected entangled pair states (PEPS) constitute the main class of tensor network states describing gapped phases \cite{Verstraete04}. In SPT phases described by PEPS, the global on-site symmetry action of $G$ is translated at the boundary to a matrix product operator (MPO) representation of $G$; $U_g U_h = U_{gh}$. This MPO, an operator with a tensor network structure, has a discrete label  $\omega \in \mathcal{H}^{3}(G,U(1))$ that characterizes the SPT phase. 

If the SPT phase is non-trivial, which is characterized by an $\omega$ that does not belong to the trivial class $[1]\in \mathcal{H}^{3}(G,U(1))$, the MPO cannot be on-site $U_g \neq \bigotimes_i u_g$. If we consider a one-dimensional system placed at the boundary which is invariant under the MPOs ,we say that the 1D system  has a symmetry characterized by the anomaly $\omega$.

\begin{figure}[h!]
 \centering
 \includegraphics[scale=1]{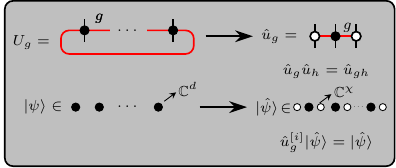}
\caption{Sketch of the procedure proposed in this paper to gauge MPO symmetries at the level of states. Left: the initial configuration with the MPO fulfilling $U_gU_h=U_{gh}$ and $U_g |\psi \rangle = |\psi \rangle$. Right: the local symmetry operators $\hat{u}_g$ are constructed by localizing the MPO $U_g$ and become the local symmetry of the state $|\hat{\psi}\rangle $ }
\label{fig:sketch}
\end{figure}

In this manuscript we explore the possibility of gauging one-dimensional states with MPO symmetries. The key is that we are able to exploit the local tensor structure of the MPOs to construct local operators from them, see Fig.\ \ref{fig:sketch}. Concretely, it has been shown \cite{Andras18B, Garre22} that under the appropriate conditions, the group relations $U_g U_h = U_{gh}$ of the MPOs are encoded locally, i.e.\ that there exist so-called {\it fusion tensors} that carry out the decomposition of the product between the local tensors of $U_g$ and $U_h$ to the local tensor of $U_{gh}$. This allows us to construct local operators $\hat{u}_g$, which satisfy $\hat{u}_g \hat{u}_h = \hat{u}_{gh}$ for any MPO, even for anomalous ones. 

For non-anomalous MPOs, $\omega \in [1]$, we are able to use such local operators to generalize the approach taken in Ref.\  \cite{Haegeman14} for gauging quantum states with on-site symmetries to MPO symmetries. This approach implements a projector to the gauge-invariant subspace by a group averaging procedure. This projector can then be applied to any matter state by first embedding it in the total gauge-and-matter Hilbert space. 

For anomalous MPOs we identify where the obstruction to the above gauging procedure appears. However, we can still provide a symmetrization procedure using the local operators that completely decouples the matter fields: it results in a trivial gauge theory. 

The manuscript is structured as follows. In Section \ref{sec:background} we cover the necessary  background we build upon: we recall the gauging procedure for on-site symmetries that we later emulate \cite{Haegeman14}; we review matrix product states (MPS) and their local on-site symmetries and finally we introduce matrix product operator (MPO) representations of symmetries and MPS symmetric under such MPOs. In Section \ref{sec:Gaugnonano} we apply the group averaging approach \cite{Haegeman14}
to non-anomalous MPO representations. We provide  gauging procedures for states and operators that are compatible. We show how the gauging of a normal subgroup $N$ of $G$ results in a global symmetry given by the quotient $G/N$. We also show that when our procedure is particularized to on-site symmetries we recover the one of Ref.\ \cite{Haegeman14}. We end the section by applying the averaging procedure to a symmetric MPS to obtain another MPS where we provide the  tensor for the gauge fields explicitly. In section \ref{sec:MPOtopo} we show how to localize anomalous MPOs and point out where the gauging procedure for non-anomalous MPOs fails in the anomalous case. 
We include a similar localization procedure for anomalous MPOs representing fusion category symmetries in Appendix \ref{sec:gauFCsym}. Finally, in Section \ref{outlook} we conclude and comment on the outlook of this work.

\section{Background: gauging on-site symmetries and tensor networks}\label{sec:background}

\subsection{Gauging quantum states}\label{sec:gauonsite}
Similar to the case of Lagrangians and Hamiltonians, gauging at the level of states should be thought of  as a map 
$\G : \H_{\mathrm{m}}\rightarrow \H_{\mathrm{m+g}}$,
which when acting on states with a global symmetry,
$\bigotimes_i u_i(g)\ket{\psi}=\ket{\psi}$
produces a state with a local symmetry
$ \hat{u}_i(g)\ket{\hat{\psi}}=\ket{\hat{\psi}}$ for all $i$. 
The target Hilbert space includes additional degrees of freedom (d.o.f.) corresponding to the gauge field $\H_{\mathrm{m}+\mathrm{g}}=\H_{\mathrm{m}}\otimes\H_{\mathrm{g}}$. 
The gauge d.o.f.\ are positioned on the edges connecting the vertices hosting the matter d.o.f. Here we are dealing with 1D spin chains, then for every site $i$ of the chain we refer to the d.o.f.\ to the left and to the right of spin $i$ as sites $[i]_l$ and $[i]_r$ respectively, with $[i]_r$ and $[i+1]_l$ denoting the same site.
The local Hilbert space around site $i$ is then 
 \[ \hat{\H}_i^{\rm local} := \H_{[i]_l} \otimes \H_i \otimes \H_{[i]_r} ,\]
where the local group representation  $\hat{u}_i$ acts. 
Each gauge d.o.f.\ at site $[i]_r\equiv[i+1]_l$ can be transformed by the action of either $\hat{u}_{i}$ or $\hat{u}_{i+1}$. The Hilbert space representing the gauge d.o.f.\ is chosen to be  $\mathbb{C}[G]$ the space of (square integrable) complex functions of the group elements with basis $\{ |g\rangle, g\in G \}$.
This space admits the left and right regular representations: $L_g|h\rangle = |gh\rangle$ and $R_g|h\rangle = |hg^{-1}\rangle$. The local group action on  $\hat{\H}_i^{\rm local}$ is then defined as  $\hat{u}^{[i]}_g=R^{[i]_l}_g\otimes u^{[i]}_g \otimes L^{[i]_r}_g$.

In Ref.\ \cite{Haegeman14}, it was shown that  the local group operators $\hat{u}_i$ can be used to construct a projector to the gauge invariant subspace. First we define the local projectors by
\[ \P_i = \frac{1}{|G|}\sum_{g\in G} \hat{u}^{[i]}_g \; 
\]
(or by the corresponding integral in the Lie group case). 
Then the global projector is defined as $\P=\Pi_i \P_i$, where the order in the product is immaterial since the left and the right group representations commute $[L_h,R_g]=0$.

One can construct gauge invariant states by taking a matter state $\ket{\psi}\in \H_{\rm m}$,  coupling it with an appropriate state of the gauge fields $\otimes_i |e\rangle_{[i]_r}$, and applying the global projector $\mathcal{P}$, such that the gauging map is $\mathcal{G} | \psi \rangle = \mathcal{P} \left(|\psi \rangle \ \bigotimes_{i}|e\rangle_{[i]_r} \right )$.

Similarly, one defines a gauging map for operators 
$\Gamma:\L(\H_{\mathrm{m}})\rightarrow \L(\H_{\mathrm{m+g}})$. 
These procedures satisfy the properties 
\begin{enumerate}
    \item $\hat{u}^{[i]}_g \G = \G$   and  $\left[\hat{u}^{[i]}_g, \Gamma[O]\right] = 0$
    \item $\G \cdot U_g =\G$,
    \item $\Gamma[O] \G\ket{\psi}=\G O\ket{\psi}$ when $[O,U_g]=0$.
\end{enumerate}

One also can gauge just part of the global symmetry, in particular a normal subgroup $N$ of $G$. It can be seen that the remaining global symmetry is given by the quotient $G/N$ realized by the operators $\hat{U}_g= \bigotimes_i (u_g \otimes R_g L_g)=\Pi_i \hat{u}^{[i]}_g$, where $R_g L_g$ is an automorphism of $N$ by conjugating with $g$: $n\rightarrow gng^{-1}$. These operators satisfy
\begin{equation*}
\hat{U}_g \G = \G U_g \ .
\end{equation*} 
The partially gauged state is therefore invariant under any operator of the form
\begin{equation}
\label{eq:remLocSymmOnsite}
\hat{U}_{[g]}:=\bigotimes_{g_i \in  [g]} (u_{g_i} \otimes R_{g_i} L_{g_i}) = \prod_{g_i\in[g]} \hat{u}_{g_i}^{[i]} \ ,
\end{equation} 
where $[g] \in G/N$ is  a coset of $G$ by $N$.

\subsection{Matrix product states and their global on-site symmetries}

MPS are vectors defined by a set of $D\times D$ matrices $A^i$, where $i=1,\dots, d$,  constructed as 
$$|\psi_A \rangle=  \tr(A^{i_1}\cdots A^{i_L})|{i_1} \cdots {i_L}\rangle \in (\mathbb{C}^d)^{\otimes L},$$
where we consider the translationally invariant case with periodic boundary conditions (PBC) for simplicity. We represent the MPS defined by the tensor $A$ graphically     as
\begin{equation*}
|\psi_{A} \rangle =
		\begin{tikzpicture}
		  \draw[rounded corners] (0.4,0) rectangle (2.6,-0.5);
		  \foreach \x in {0.75,2.25}{
		    \node[tensor,label=below:$A$] (t\x) at (\x,0) {};
		    \draw (\x,0) --++ (0,0.3);
		  }
		  \node[fill=white] at (1.5,0) {$\dots$};
		\end{tikzpicture} \ .
	\end{equation*}

An MPS tensor is called {\it injective} if it is injective  as a map from the virtual space to the physical \cite{PerezGarcia07}. An injective MPS is the unique ground state of its so-called parent Hamiltonian. Moreover, MPSs are in  the family of tensor network states describing the ground states of 1D local gapped Hamiltonians \cite{Hastings06,Hastings07A,Hastings07B}.

We say that the MPS is symmetric under the global on-site operator $U= u^{\otimes L}$ if $U |\psi_{A} \rangle = |\psi_{A} \rangle$. If the tensor $A$ is injective, the symmetry under $U$ can be characterized locally \cite{Molnar18A}. Namely, there exists a (projective) representation $V:G\rightarrow \mathcal{M}_D$ such that the following holds for all $g\in G$:
\begin{equation}\label{eq:onsym}
		\begin{tikzpicture}
		    \node[tensor,label=below:$A$] (t) at (0,0) {};
		    \node[tensor,label=right:$u_g$,scale=0.7]  at (0,0.3) {};
		    \draw (0,0) -- (0,0.5);
		     \draw (-0.3,0) -- (0.3,0);
		\end{tikzpicture}
		=
		\begin{tikzpicture}
		    \node[tensor,label=below:$A$] (t) at (0,0) {};
		    \node[tensor,label=above:$V_g^\dagger$,scale=0.7]  at (0.3,0) {};
		    \node[tensor,label=above:$V_g$,scale=0.7]  at (-0.3,0) {};
		    \draw (0,0) -- (0,0.5);
		     \draw (-0.45,0) -- (0.45,0);
		\end{tikzpicture}\ .
	\end{equation}

\subsection{Matrix product operator symmetries} \label{sec:MPOsymMPS}

An MPO is constructed by a set of $d^2$  matrices $\{T ^{i,j}\}_{i,j=1,\cdots, d}$ of dimensions $\chi \times \chi$, where $\chi$ is called the bond dimension of the MPO and $i,j=1,\cdots, d$ label a basis of the local Hilbert space $\mathcal{H} =\mathbb{C}^d$. The MPO generated by the tensor $T$ on a chain of length $L$ with PBCs is:
$$ O_T = \sum_{ \{ i_k\} }\tr(T^{i_1,j_2}\cdots T^{i_L,j_L} )|i_1\cdots i_L \rangle \langle  j_1 \cdots j_L|\ .$$
In this section we consider MPO representations of a finite group $G$ in a subspace. We denote the MPO corresponding to the element $g\in G$ by $U_g= O_{T_g}$, where the tensor $T_g$ is injective with bond dimension $\chi_g$. The MPO $U_g$ is represented graphically as: 
\begin{equation*}
U_g =
			\begin{tikzpicture}
		  \draw[red,rounded corners] (0.25,0.3) rectangle (2.75,-0.2);
		  \foreach \x in {0.75,2.25}{
		    \node[tensor] (t\x) at (\x,0.3) {};
		    \draw (\x,0) --++ (0,0.6);
		    \node[] at (\x+0.25,0.5) {$T_g$};
      }
		  \node[fill=white] at (1.5,0.3) {$\dots$};
		\end{tikzpicture}   \ .
	\end{equation*}
These MPOs satisfy $U_g U_h = U_{gh}$ and we assume that there are fusion tensors 
$$
W_{g,h} =
\begin{tikzpicture}
\pic  at (0,0) {W1=gh/g/h};
\end{tikzpicture} 
\ , \
W^{-1}_{g,h} =
\begin{tikzpicture}
\pic  at (0,0) {V1=gh/g/h};
\end{tikzpicture} 
$$
that satisfy the following:
\begin{equation}\label{eq:fusiontensors}
  \begin{tikzpicture}
    \draw[red]  (-0.5,0)--(0.5,0);
    \draw[red]  (-0.5,-0.5)--(0.5,-0.5);
    \node[irrep] at (-0.35,0) {$g$};
    \node[irrep] at (-0.35,-0.5) {$h$};
    \node[tensor] (t) at (0,0) {};
    \node[tensor] (t) at (0,-0.5) {};
    \draw (0,-0.8) -- (0,0.3);
  \end{tikzpicture} =
  \begin{tikzpicture}[baseline=-1mm]
    \node[tensor] (t) at (0,0) {};
    \draw (0,-0.3) -- (0,0.3);
    \pic (v) at (0.5,0) {V1=gh/g/h};
    \pic (w) at (-0.5,0) {W1=gh/g/h};
  \end{tikzpicture}\ ,
\end{equation}
together with the orthogonality relations:
\begin{equation}\label{orthoMPOG}
   \begin{tikzpicture}
    \pic (v) at (0,0) {V1=gh/g/h};
    \pic (w) at (0.6,0) {W1=k//};
   \end{tikzpicture} 
= \delta_{gh,k} \id_{gh} \ .
\end{equation}
The MPOs are a representation of $G$ on the subspace $U_e {\cdot } \bigotimes_i\mathcal{H}_i^{\rm local}$  since $U_e$ is not in general the identity (it is the projector onto the invariant subspace).
The associativity of the MPO product implies that the fusion tensors are associative up to a phase factor:
\begin{equation}\label{eq:3cocycle}
  \begin{tikzpicture}
    \pic (w1) at (0,0) {V1=/gh/k};
    \node[irrep, anchor = south] at (w1-mid) {$ghk$};
    \pic (w2) at (w1-up) {V1=/g/h};
  \end{tikzpicture} 
  = \omega(g,h,k)
  \begin{tikzpicture}
    \pic (w1) at (0,0) {V1=/g/hk};
    \node[irrep, anchor = south] at (w1-mid) {$ghk$};
    \pic (w2) at (w1-down) {V1=/h/k};
   \end{tikzpicture} \ ,
\end{equation}
where $\omega \in U(1)$ is a $3$-cocycle since it satisfies
$$ \omega(g, h, k) \omega(g, h k,l) \omega(h, k, l) = \omega(g h, k, l)\omega(g, h, k l).$$

Eq.\ \eqref{eq:fusiontensors} is invariant under $W_{g,h}\rightarrow \beta_{g,h} W_{g,h} $ and $\myinv{W}_{g,h}\rightarrow \bar{\beta}_{g,h} \myinv{W}_{g,h}$ where $\beta \in U(1)$. That change modifies the $3$-cocycle as $\omega \rightarrow  \omega \frac{\beta_{h,k}\beta_{g,hk}}{\beta_{g,h}\beta_{gh,k}}$: multiplication by a $3$-coboundary. Then, $3$-cocycles are classified by quotienting by these transformations, which results in the third cohomology group of $G$: $\mathcal{H}^3(G,U(1))$.

In what follows we consider an injective MPS $ |\psi_{A} \rangle $, with bond dimension $D$, that is invariant under the MPOs:
\begin{equation}\label{symcond}
U_g |\psi_{A} \rangle = |\psi_{A} \rangle \ , \forall g\in G \ .
\end{equation}

We assume that the action of the MPOs on the MPS is realized locally by a set of action tensors 
\begin{equation}\label{actensor}
V_{g} =
\begin{tikzpicture}
\pic  at (0,0) {W2=/g/};
\end{tikzpicture} \ ,
\end{equation}
that satisfy:
\begin{equation}\label{eq:actiontensors}
  \begin{tikzpicture}
    \draw[red] (-0.5,0)--(0.5,0);
     \node[irrep] at (0.55,0) {$g$};
    \draw (-0.5,-0.5)--(0.5,-0.5);
    \node[tensor] (t) at (0,0) {};
    \node[tensor,label=below:$A$] (t) at (0,-0.5) {};
    \draw (0,-0.5) -- (0,0.3);
  \end{tikzpicture} 
\ =  \ 
  \begin{tikzpicture}
    \pic[] (v) at (0.5,0) {V2=/g/};
    \pic[] (w) at (-0.5,0) {W2=/g/};
    \node[tensor,label=below:$A$] (t) at (0,0) {};
    \draw (0,0) -- (0,0.3);
  \end{tikzpicture} 
\end{equation}
and the orthogonality relations:
\begin{equation}\label{eq:orthoacten}
   \begin{tikzpicture}
    \pic (v) at (0,0) {V2=/g/};
    \pic (w) at (0.6,0) {W2=//};
   \end{tikzpicture} 
   = \  \id_D \ .
\end{equation}
Associativity of the MPO action on the MPS results in the following equation:

\begin{equation}\label{eq:Ldef}
  \begin{tikzpicture}
    \pic[] (v1) at (1,0) {V2=/gh/};
    \pic (v2) at (v1-up) {V1=/g/h};
  \end{tikzpicture} 
  = L_{g,h}
  \begin{tikzpicture}
    \pic (v1) at (1,0) {V2=/g/};
    \pic[] (v2) at (v1-down) {V2=/h/};
  \end{tikzpicture} \ .
\end{equation}
Assuming Eq.\ \eqref{symcond} already constrains the possible MPOs: $\omega$ should belong to the trivial class. For on-site symmetries, $L_{g,h}$ is the $2$-cocycle that classifies SPT phases \cite{Schuch11,Chen11}. For a complete analysis see Ref.\  \cite{Garre22}, where it is shown that the different classes of $L$-symbols, previously defined, classify the quantum phases of MPSs protected by MPO symmetries.


\section{Gauging non-anomalous MPO symmetries}\label{sec:Gaugnonano}

In this section we propose a procedure to gauge non-anomalous MPO symmetries, which are representations of groups when they are restricted to a certain subspace. These MPOs arise naturally as symmetries in the boundaries of 2D both trivial intrinsic and trivial symmetry protected topologically ordered states described by projected entangle pair states \cite{Schuch10,Sahinoglu14,Chen11A,Andras18B}.

\subsection{MPO properties}

In this section we consider an MPO representation of a finite group $G$ satisfying  Eqs.\  \eqref{eq:fusiontensors} and  \eqref{orthoMPOG}.
We further require the   existence of the following:
\begin{itemize}
\item A vector $|v\rangle\in \mathbb{C}^{\chi_e}$ 
satisfying 

\begin{equation}\label{eq:vectore}
  \begin{tikzpicture}
      \pic (w) at (0,-0.3) {W1={g}/g/{e}};
    \node[tensor,label= west:$\langle v |$] (t) at (w-down) {};
  \end{tikzpicture}
=
  \begin{tikzpicture}
    \draw[red] (-0.3,0) -- (0.3,0);
   \node[irrep] at (0,0.0) {${g}$};
  \end{tikzpicture}
  =
     \begin{tikzpicture}
      \pic (w) at (0,-0.3) {V1={g}/g/{e}};
    \node[tensor,label= east:$| v \rangle $] (t) at (w-down) {};
  \end{tikzpicture} \ ,
\end{equation}
where the equations also hold when $|v\rangle$ is applied on the upper legs,

\item a matrix $Z_g$ for every group element that maps the block $g$ to $g^{-1}$ and satisfies

\begin{equation}\label{eq:Zmatrix}
 \begin{tikzpicture}
    \pic (v2) at (-0.25,0) {W1=h/g/};
    \node[irrep] at (-0.6,-0.3) {$\myinv{g}h$};
    \end{tikzpicture}
    =
    \begin{tikzpicture}
    \pic (v1) at (0,0) {V1=//h};
    \node[irrep] at (-0.4,0.0) {$\myinv{g}h$};
    \draw[rounded corners,red] (v1-up)--++ (0.15,0)--++(0,0.35)--++ (-0.6,0);
    \draw[fill=black] (0.6,0.43) rectangle (0.6+0.1,0.43+0.1);
    \node[irrep] at (0.3,0.65) {$g$};
    \node[irrep] at (0.3,0.3) {$\myinv{g}$};
    \node[] at (0.9,0.45) {$Z_g$};
    \end{tikzpicture} \ .
\end{equation}

\end{itemize}
We remark that these conditions are satisfied for the regular 'triple line' representation of MPOs \cite{Bultinck17A} and they have been proven for certain algebra structures in Ref.\ \cite{molnar22}. We notice that combining  Eqs.\ \eqref{eq:vectore} and \eqref{eq:Zmatrix} we obtain the defining property of $Z_g$:
\begin{equation}\label{eq:Zmatrixdef}
 \begin{tikzpicture}
    \pic (v2) at (-0.25,0) {W1=e \ /g/};
    \node[tensor,label=right:$|v \rangle$] (t) at (0.15,0) {};
    \node[irrep] at (-0.6,-0.3) {$\myinv{g}$};
    \end{tikzpicture}
    =
    \begin{tikzpicture}
    \draw[rounded corners,red] (0,0.3)--(0.4,0.3)-- (0.4,-0.30)--(0,-0.30);
    \draw[fill=black] (0.4-0.05,0-0.05) rectangle (0.4+0.1,0+0.1);
    \node[irrep] at (0.1,0.3) {$g$};
    \node[irrep] at (0.1,-0.3) {$\myinv{g}$};
    \node[] at (0.7,0) {$Z_g$};
    \end{tikzpicture} \ .
\end{equation}

The non-anomalous condition comes from the fact that the $3$-cocycle of the fusion tensors is trivial, i.e. $\omega  \beta_{h,k} \beta_{g,hk} \myinv{\beta_{g,h}} \myinv{\beta_{gh,k}}=1$ it is a $3$-coboundary. This means that there is a transformation on the fusion tensors, namely multiplication by the phase factor $\beta_{g,h}$, that makes $\omega = 1$ for all group elements such that they satisfy 

\begin{equation}\label{3cocytriv} 
  \begin{tikzpicture}
    \pic (w1) at (0,0) {V1=/gh/k};
    \node[irrep, anchor = south] at (w1-mid) {$ghk$};
    \pic (w2) at (w1-up) {V1=/g/h};
  \end{tikzpicture} =
  \begin{tikzpicture}
    \pic (w1) at (0,0) {V1=/g/hk};
    \node[irrep, anchor = south] at (w1-mid) {$ghk$};
    \pic (w2) at (w1-down) {V1=/h/k};
   \end{tikzpicture} \ ,
\end{equation}

and using Eq.\ \eqref{orthoMPOG} we arrive at

\begin{equation}\label{eq:trivial3ortho}
\begin{tikzpicture}
    \pic (v1) at (-1.5,0) {V1=ghk \ \ /g/hk};
    \pic (v2) at (v1-down) {V1=/h/k};
    \pic[scale=0.7] (w) at (-0.5,0.21) {W1=\ gh//};
    \draw[red] (v1-up) to [out=0,in=180] (w-up);
    \end{tikzpicture}
    =
\begin{tikzpicture}
        \pic (w2) at (w1-down) {V1=ghk \ \ /\ gh/k};
   \end{tikzpicture} \ ,
\end{equation}
which will allow us to properly localized the MPO.

\subsection{Gauging procedure}
We assume that there is a state $|\psi \rangle \in \mathcal{H}^{\rm m}=(\mathbb{C}^d)^{\otimes L}$ invariant under the action of MPOs:
$$U_g|\psi \rangle = |\psi \rangle,$$
for every $g\in G$.

As in Section \ref{sec:gauonsite}, the gauging procedure introduces gauge fields by decorating each edge   of the chain with a Hilbert space $\bigoplus_g \mathbb{C}^{\chi_g}$, where $\chi_g$ corresponds to the MPO bond dimension of the block $g$. As above, we refer to the edge to the left (right) of site $i$ by $[i]_l$($[i]_r$) and identify $[i]_r=[i+1]_l$.

We define the local operators acting on $\hat{\mathcal{H}}_i^{\rm local}=\H_{[i]_l}\otimes \H_i \otimes \H_{[i]_r}$, which will correspond to the local symmetry of the gauged state, using the MPO tensor and the fusion tensors as follows: 
\begin{equation}\label{eq:locsym}
\hat{u}_g =\sum_{h,k\in G}
  \begin{tikzpicture}
     \pic[scale=0.8] (v) at (-0.6,0.15) {V1p=//g};
      \pic[scale=0.8] (w) at (0.6,-0.15) {W1p=/g/k};
       \node[irrep] at (-0.6,0.5) {$h\myinv{g}$};
      \node[irrep] at (1,0.2) {${gk}$};
    \node[irrep] at (-0.8,-0.6) {${h}$};
    \draw[rounded corners,red] (v-mid)--++ (-0.15,0)--++(0,-0.4);
    \draw[rounded corners,red] (v-up)--++(0.1,0)--++(0,0.4);
  \draw[rounded corners,red] (w-down)--++(-0.1,0)--++(0,-0.2);
    \draw[rounded corners,red] (w-mid)--++ (0.15,0)--++(0,0.4);
     \draw[red] (v-down) to [out=0,in=180] (w-up);
    \node[tensor] (t) at (0,0) {};
    \draw (t)-++ (0,0.3)-++ (0,-0.3);
  \end{tikzpicture} \ .
\end{equation}

Let us first show that the operators so defined satisfy the group property $\hat{u}_{g_1}\hat{u}_{g_2}=\hat{u}_{g_1g_2}$:

\begin{align*}
\hat{u}_{g_1}\hat{u}_{g_2} = & \sum_{h,k}
    \begin{tikzpicture}[baseline=-3mm]
    \pic (v1) at (-1.2,0.15) {V1m=//g_2};
    \pic (v2) at (v1-up) {V1u=//g_1};
    \draw[rounded corners,red] (v1-mid)--++ (-0.15,0)--++(0,-0.5);
    \node[irrep] at (-1.4,-0.6) {${h}$};
    \draw[rounded corners,red] (v2-up)--++(0.1,0)--++(0,0.3);
    \pic (w1) at (1.2,-0.15) {W1m=/g_1/};
    \pic (w2) at (w1-down) {W1D=/g_2/k};
    \draw[rounded corners,red] (w1-mid)--++ (0.15,0)--++(0,0.5);
    \draw[rounded corners,red] (w2-down)--++(-0.1,0)--++(0,-0.3);
    \draw[red] (v2-down)-- (w1-up);
    \draw[red] (v1-down)-- (w2-up);
    \node[tensor] (t1) at (0,-0.15) {};
    \node[tensor] (t2) at (0,0.15) {};
    \draw (0,0.45)--(0,-0.45);
   \end{tikzpicture} 
   \\ = & \sum_{h,k}
       \begin{tikzpicture}[baseline=-3mm]
    \pic (v1) at (-1.5,0.15) {V1m=//g_2};
    \pic (v2) at (v1-up) {V1u=//g_1};
        \draw[rounded corners,red] (v1-mid)--++ (-0.15,0)--++(0,-0.5);
    \draw[rounded corners,red] (v2-up)--++(0.1,0)--++(0,0.3);
    \pic[scale=0.7] (w3) at (-0.5,-0.06) {W1=//};
    \draw[red] (v1-down) to [out=0,in=180] (w3-down);
    \node[irrep] at (-1.7,-0.6) {${h}$};
    \pic (w1) at (1.5,-0.15) {W1m=/g_1/};
    \pic (w2) at (w1-down) {W1D=/g_2/k};
        \draw[rounded corners,red] (w1-mid)--++ (0.15,0)--++(0,0.5);
    \draw[rounded corners,red] (w2-down)--++(-0.1,0)--++(0,-0.3);
        \pic[scale=0.7] (v3) at (0.5,0.06) {V1=//};
    \draw[red] (w1-up) to [out=180,in=0] (v3-up);
    \draw[red] (v3-mid) to [out=180,in=0] (w3-mid);
    \node[tensor] (t1) at (0,0) {};
    \draw (0,0.45)--(0,-0.45);
   \end{tikzpicture} 
    \\ = & \sum_{h,k}
  \begin{tikzpicture}
     \pic[scale=0.8] (v) at (-0.6,0.15) {V1p=//};
      \pic[scale=0.8] (w) at (0.6,-0.15) {W1p=//k};
       \node[irrep] at (-0.65,0.5) {$h\myinv{g_{12}}$};
      \node[irrep] at (0.9,0.25) {${g_{12}k}$};
    \node[irrep] at (-0.8,-0.6) {${h}$};
    \draw[rounded corners,red] (v-mid)--++ (-0.15,0)--++(0,-0.4);
    \draw[rounded corners,red] (v-up)--++(0.1,0)--++(0,0.4);
  \draw[rounded corners,red] (w-down)--++(-0.1,0)--++(0,-0.2);
    \draw[rounded corners,red] (w-mid)--++ (0.15,0)--++(0,0.4);
     \draw[red] (v-down) to [out=0,in=180] (w-up);
     \node[irrep] at (-0.25,0) {$g_{12}$};
    \node[tensor] (t) at (0,0) {};
    \draw (t)-++ (0,0.3)-++ (0,-0.3);
  \end{tikzpicture}
  = \hat{u}_{g_1g_2} \ , 
\end{align*}
where we denote $g_{12}:=g_1g_2$, and used Eq.\ \eqref{eq:fusiontensors} in the first equality and Eq.\ \eqref{eq:trivial3ortho} in the second one.

The operators $\hat{u}^{[i]}_{g}$ and $\hat{u}^{[i+1]}_{g'}$ overlap because both act on $[i]_r= [i+1]_l$. In the case of a non-anomalous symmetry the neighboring operators commute. We now show that
\begin{equation}\label{eq:commU}
     [ \hat{u}^{[i]}_{g} ,\hat{u}^{[i+1]}_{g'}] =0 
    \ .
\end{equation} 
To do so it is enough to prove Eq.\ \eqref{eq:commU} in the Hilbert space where the operators overlap, i.e.\  $\H_{[i]_r}= \bigoplus_g \mathbb{C}^{\chi_g} $, and in every block $\mathbb{C}^{\chi_h}$ of the total direct sum, so we need to prove that for all $h,g$ and $g^\prime$
\begin{equation}\label{eq:commUsimp}
  \begin{tikzpicture}
      \pic[scale=0.8] (w) at (0.2,0.45) {W1p=/g \ \ /};
      \draw[red] (w-up) --++ (-1,0);
            \node[tensor] at (-0.5,0.7) {};
            \draw (-0.5,0.4)--(-0.5,1);
 \pic[scale=0.8] (v) at (-0.2,-0.45) {V1p=h//\ \ g'};
      \draw[red] (v-down) --++ (1,0);
            \node[tensor] at (0.5,-0.7) {};
            \draw (0.5,-0.4)--(0.5,-1);
      \draw[red] (v-up) to [out=0,in=180] (w-down);
    \draw[rounded corners,red] (v-mid)--++ (-0.15,0)--++(0,-0.4);
    \draw[rounded corners,red] (w-mid)--++ (0.15,0)--++(0,0.4);
  \end{tikzpicture}
=
    \begin{tikzpicture}
     \pic[scale=0.8] (v) at (0.2,0.3) {V1p=//\ \ g'};
           \draw[red] (v-down) --++ (0.8,0);
            \node[tensor] at (0.7,0.05) {};
            \draw (0.7,-0.25)--(0.7,0.35);
      \pic[scale=0.8] (w) at (-0.2,-0.3) {W1p=/g/h};
      \draw[red] (w-up) --++ (-0.8,0);
            \node[tensor] at (-0.7,-0.05) {};
            \draw (-0.7,-0.35)--(-0.7,0.25);
      \draw[red] (v-mid) to [out=180,in=0] (w-mid);
    \draw[rounded corners,red] (w-down)--++ (-0.15,0)--++(0,-0.4);
    \draw[rounded corners,red] (v-up)--++ (0.15,0)--++(0,0.4);
  \end{tikzpicture} \ .
\end{equation}
The LHS of Eq.\ \eqref{eq:commUsimp}, without the MPO tensors, can be rewritten as
$$ 
\begin{tikzpicture}
    \pic (v1) at (0,0) {V1=h//g'};
    \pic (v2) at (0.6,0.6) {W1=/g/};
    \end{tikzpicture}
    =
    \begin{tikzpicture}
    \pic (v1) at (0,0) {V1=h//g'};
    \pic (v2) at (v1-up) {V1=/\myinv{g}/};
        \node[irrep] at (0.7,0.9) {$g$};
    \draw[rounded corners,red] (v2-up)--++ (0.15,0)--++(0,0.35)--++ (-0.6,0);
    \draw[fill=black] (1.1,0.73) rectangle (1.1+0.1,0.73+0.1);
    \node[] at (1.4,0.75) {$Z_g$};
    \end{tikzpicture}
    $$
by using the property Eq.\ \eqref{eq:Zmatrix}. Also, using again  the property Eq.\ \eqref{eq:Zmatrix} the RHS of Eq.\ \eqref{eq:commUsimp} is
$$ 
\begin{tikzpicture}
    \pic (v1) at (0.25,0) {V1=//g'};
    \pic (v2) at (-0.25,0) {W1=/g/h};
    \end{tikzpicture}
    =
    \begin{tikzpicture}
    \pic (v1) at (0,0) {V1=h/\myinv{g}/};
    \node[irrep] at (0.3,0.65) {$g$};
    \pic (v2) at (v1-down) {V1=//g'};
    \draw[rounded corners,red] (v1-up)--++ (0.15,0)--++(0,0.35)--++ (-0.6,0);
    \draw[fill=black] (0.6,0.43) rectangle (0.6+0.1,0.43+0.1);
    \node[] at (0.9,0.5) {$Z_g$};
    \end{tikzpicture} \ .
    $$
Then, applying $\myinv{Z}_g$ to both sides and using Eq.\ \eqref{3cocytriv} proves Eq.\ \eqref{eq:commUsimp} and subsequently Eq.\ \eqref{eq:commU}.

With the desired commutation relation at hand we can define the local projectors:
$$\mathcal{P}^{[i]} = \frac{1}{|G|}\sum_g \hat{u}^{[i]}_g, $$
so that they satisfy $\mathcal{P}^2 = \mathcal{P}$ and also
$[\mathcal{P}^{[i]} ,\mathcal{P}^{[j]} ]=0$ because of Eq.\ \eqref{eq:commU}. The total projector onto the gauge invariant subspace is then $\mathcal{P} = \prod_{i} \mathcal{P}^{[i]}$.

Finally the gauging procedure $\mathcal{G}: \mathcal{H}^{\rm m} \rightarrow \mathcal{H}^{\rm m} \otimes \mathcal{H}^{\rm g}$ maps a  state $|\psi\rangle \in \mathcal{H}^{\rm m} $ symmetric under $U_g$ to a state symmetric under $\hat{u}^{[i]}_{g}$ for all $g\in G $  and for all $i$. The procedure is obtained by first coupling $\ket{\psi}$ to a certain state of the gauge field and then applying the projector to the invariant subspace:
$$\G | \psi \rangle = \mathcal{P} \left(|\psi \rangle \ \bigotimes_{i}|v\rangle_{[i]_r} \right ) ,$$
where $|v\rangle$  is the vector supported on the unit block $e$ of the MPO satisfying Eq.\ \eqref{eq:vectore}.

Using Eq.\ \eqref{eq:vectore} we can write the action of the  gauging map on the matter Hilbert space
explicitly:
\begin{equation}\label{eq:gaugeop}
\G    =
\sum_{\{g_i\}}\ 
\begin{tikzpicture}[baseline=-1]
\node[irrep] at (-0.6,0) {$\cdots$};
\node[irrep] at (4.2,0) {$\cdots$};
\draw[red] (-0.3,0) -- (0,0); 
\foreach \x in {0,1.3,2.6}{
\pic[scale=0.6] (v) at  ($ (\x,0) + (0.5,0) $) {V1p=//};
\draw[red] (v-mid) -- (\x,0); 
    \node[tensor] (t) at (\x,0) {};
    \draw (t)-++ (0,0.3)-++ (0,-0.3);
\draw[red,rounded corners] (v-up)--++(0.125,0)--++(0,0.25);
\draw[red] (v-down) to [out=0,in=180] ($(\x,0) + (1.3,0)$);
}
 \node[irrep] at (0.2,0) {$g_i$};
 \node[irrep] at (0.9,-0.1) {$g_{\scriptstyle i+1}$};
 \node[irrep] at (2.2,-0.1) {$g_{\scriptstyle i+2}$};
  \node[irrep] at (3.5,-0.1) {$g_{\scriptstyle i+3}$};
   \node[irrep] at (0.6,0.4) {$g_i\myinv{g}_{\scriptstyle i+1}$};
    \node[irrep] at (1.9,0.4) {$g_{\scriptstyle i+1}\myinv{g}_{\scriptstyle i+2}$};
    \node[irrep] at (3.3,0.4) {$g_{\scriptstyle i+2}\myinv{g}_{\scriptstyle i+3}$};
\end{tikzpicture} \ .
\end{equation}

In what follows we show that
\begin{equation}\label{MPOgaugedinv}
 \mathcal{G}  U_g  = \mathcal{G}  \ . 
\end{equation}
To do so we write explicitly the following,
$$
\G {U}_{g} =  \sum_{\{g_i\}}\ 
\begin{tikzpicture}
\node[] at (-0.5,-0.2) {$\cdots$};
\node[] at (3.1,-0.2) {$\cdots$};
\draw[red] (-0.3,0) -- (0,0); 
\draw[red] (2.6,0) -- (2.9,0); 
\draw[red] (-0.3,-0.4) -- (3,-0.4);
\node[irrep] at (0.7,-0.65) {$g$};
\foreach \x in {0,1.3}{
\pic[scale=0.6] (v) at  ($ (\x,0) + (0.5,0) $) {V1p=//};
\draw[red] (v-mid) -- (\x,0); 
    \node[tensor] (t) at (\x,0) {};
    \draw (t)-++ (0,0.3)-++ (0,-0.7);
    \node[tensor] at (\x,-0.4) {};
\draw[red,rounded corners] (v-up)--++(0.125,0)--++(0,0.25);
\draw[red] (v-down) to [out=0,in=180] ($(\x,0) + (1.3,0)$);
}
\draw (2.6,0.3)--(2.6,-0.7);
\node[tensor] at (2.6,0) {};
\node[tensor] at (2.6,-0.4) {};
 \node[irrep] at (0.2,0) {$g_i$};
 \node[irrep] at (0.9,-0.1) {$g_{\scriptstyle i+1}$};
 \node[irrep] at (2.2,-0.1) {$g_{\scriptstyle i+2}$};
   \node[irrep] at (0.6,0.4) {$g_i\myinv{g}_{\scriptstyle i+1}$};
    \node[irrep] at (1.9,0.4) {$g_{\scriptstyle i+1}\myinv{g}_{\scriptstyle i+2}$};
\end{tikzpicture}
$$
We now can use Eqs.\ \eqref{eq:fusiontensors} and \eqref{3cocytriv} to manipulate the previous operator and we obtain
\begin{align}\label{eq:calc1}
\begin{tikzpicture}[baseline=-2mm]
\draw[red] (-0.3,0) -- (0,0); 
\draw[red] (1.3,0) -- (1.6,0); 
\draw[red] (-0.3,-0.4) -- (1.6,-0.4);
\node[irrep] at (0.7,-0.65) {$g$};
\foreach \x in {0}{
\pic[scale=0.6] (v) at  ($ (\x,0) + (0.5,0) $) {V1p=//};
\draw[red] (v-mid) -- (\x,0); 
    \node[tensor] (t) at (\x,0) {};
    \draw (t)-++ (0,0.3)-++ (0,-0.7);
    \node[tensor] at (\x,-0.4) {};
\draw[red,rounded corners] (v-up)--++(0.125,0)--++(0,0.25);
\draw[red] (v-down) to [out=0,in=180] ($(\x,0) + (1.3,0)$);
}
\draw (1.3,0.3)--(1.3,-0.7);
\node[tensor] at (1.3,0) {};
\node[tensor] at (1.3,-0.4) {};
 \node[irrep] at (0.2,0) {$g_i$};
 \node[irrep] at (0.9,-0.1) {$g_{\scriptstyle i+1}$};
   \node[irrep] at (0.6,0.4) {$g_i\myinv{g}_{\scriptstyle i+1}$};
\end{tikzpicture}
 = \ \ & 
\begin{tikzpicture}[baseline=-2mm]
\foreach \x in {0}{
\pic[scale=0.6] (v) at  ($ (\x,0) + (0.5,0) $) {V1p=//};
\draw[red,rounded corners] (v-up)--++(0.125,0)--++(0,0.25);
}
\pic[scale=0.8] (v1) at (0,-0.25)  {V1=//};
\pic[scale=0.8] (w1) at (1.3,-0.25)  {W1=//};
\draw[red] (v1-down) -- (w1-down);
\draw[red] (v-down) to [out=0,in=180] (w1-up);
\node[irrep] at (0.7,-0.5) {$g$};
\pic[scale=0.8] (v1) at (1.9,-0.25)  {V1=/g_{\scriptstyle i+1}/g};
\pic[scale=0.8] (w1) at (-0.6,-0.25)  {W1=/g_i/g};
\node[tensor] at (-0.3,-0.25) {};
\node[tensor] at (1.6,-0.25) {};
\draw (-0.3,0)--(-0.3,-0.5);
\draw (1.6,0)--(1.6,-0.5);
 \node[irrep] at (0.2,0) {$g_i$};
 \node[irrep] at (0.9,-0.1) {$g_{\scriptstyle i+1}$};
   \node[irrep] at (0.6,0.4) {$g_i\myinv{g}_{\scriptstyle i+1}$};
\end{tikzpicture}
\\ = \ \ &
\begin{tikzpicture}[baseline=-1mm]
\pic[scale=0.8] (v1) at (2,0)  {V1=/g_{\scriptstyle i+1}/g};
\pic[scale=0.8] (w1) at (-0.3,0)  {W1=/g_i/g};
\foreach \x in {0}{
\pic[scale=0.6] (v) at  ($ (\x,0) + (0.7,0) $) {V1p=//};
\draw[red] (v-mid) -- (\x,0); 
    \node[tensor] (t) at (\x,0) {};
    \draw (t)-++ (0,0.3)-++ (0,-0.3);
\draw[red,rounded corners] (v-up)--++(0.125,0)--++(0,0.25);
\draw[red] (v-down) to [out=0,in=180] ($(\x,0) + (1.6,0)$);
}
\draw (1.7,0.3)--(1.7,-0.3);
\node[tensor] at (1.7,0) {};
 \node[irrep] at (0.35,-0.05) {$g_ig$};
 \node[irrep] at (1.2,-0.1) {$g_{\scriptstyle i+1}g$};
   \node[irrep] at (0.7,0.4) {$g_i\myinv{g}_{\scriptstyle i+1}$};
\end{tikzpicture} \ , \notag
\end{align}
which shows that the action of $U_g$ on $\G$ is given by $g_i\to g_i g$, which leaves the product $g_i\myinv{g}_{\scriptstyle i+1}$ invariant  and it can be absorbed into a relabeling $g_i g\to \tilde{g}_i$ in the sum so that Eq.\ \eqref{MPOgaugedinv} is satisfied.

Therefore, if $| \Psi \rangle$ does not transform trivially under the MPOs $U_g$, {\it i.e.} $U_g | \Psi \rangle = \beta_g | \Psi \rangle$, the gauging procedure annihilates $| \Psi \rangle$ (as in the on-site case \cite{Haegeman14}). This can be seen by the characters orthogonality relations: $ \mathcal{G} | \Psi \rangle = \frac{1}{|G|}\sum_g \mathcal{G} U_g | \Psi \rangle = \frac{1}{|G|}(\sum_g \beta_g)  \mathcal{G}|\Psi \rangle = 0 $.\\

For completeness we provide a gauging procedure for operators  $\Gamma:\mathcal{L}[ \mathcal{H}^{\rm m}] \to \mathcal{L}[\mathcal{H}^{\rm m} \otimes \mathcal{H}^{\rm g}]$ such that $[\Gamma[O],\hat{u}^{[i]}_g]=0$ for any operator $O\in \mathcal{L}[ \mathcal{H}_{\rm m}]$, any site $i$ and any element $g\in G$. Let $O$ be an operator supported in some segment $\Lambda$ of the spin chain. We define the gauging procedure for operators as $\Gamma[O] = S \circ \mathcal{E}_\Lambda[O] $.
The enlarging map $\mathcal{E}_\Lambda : \mathcal{L}[ \mathcal{H}^{\rm m}_\Lambda]  \rightarrow \mathcal{L}[\mathcal{H}^{\rm m}_\Lambda \otimes \mathcal{H}^{\rm g}_\Lambda]$ is defined by

$$\mathcal{E}_\Lambda[O]= O \otimes \rho_v \ , $$ 
where $\rho_v =\bigotimes_{i\in \Lambda} |v\rangle_{[i]_r} \langle v|_{[i]_r} $.
The symmetrization map $S: \mathcal{L}[\mathcal{H}^{\rm m}_\Lambda \otimes \mathcal{H}^{\rm g}_\Lambda]  \rightarrow \mathcal{L}[\mathcal{H}^{\rm m}_\Lambda \otimes \mathcal{H}^{\rm g}_\Lambda]$ is defined as
$$S[O] =  \frac{1}{|G|}\prod_{i\in \Lambda}\sum_{g_i\in G} \hat{u}^{[i]}_{g_i} O \hat{u}^{[i]}_{g_i^{-1}} \ . $$

The gauging procedure for operators and the gauging procedure for states are compatible, that is,
\begin{equation}\label{compgau}
 \Gamma[O] \mathcal{G} | \psi \rangle = \mathcal{G} O | \psi \rangle \ ,
\end{equation}
for an operator $O$ that commutes with the MPO $U_g$ for all $g\in G$. We show this in Appendix \ref{sec:proofs}.

The gauging procedure is well defined in the sense that by projecting back to the initial gauge field configuration we recover the initial matter operators and states. By using Eq.\ \eqref{eq:vectore} we see the following for an MPO-symmetric state $| \psi \rangle$ and an operator $O$ commuting with the MPO:
$$\left ( \bigotimes_{i}\langle v|_{[i]_r} \right )  \G | \psi \rangle \propto | \psi \rangle ,\ \  \tr_{\mathrm{g}}(\rho_v \Gamma[O] ) \propto O.$$
Note that if the above relations hold with an equality instead of a proportionality relation, one can then show that the gauged state $\G | \psi \rangle$ must be a product state between gauge and matter fields.

\subsection{Gauging a subgroup and its remaining global symmetry}

In this section we consider the gauging of a normal subgroup $N$ of $G$ of a state $|\psi \rangle$ with non-anomalous MPO symmetry $\{ {U}_g , g\in G\}$. This is done by using the operators $\{ \hat{u}_n , n\in N\} $ defined as 
\begin{equation*}
\hat{u}_n =\sum_{m,\tilde{m}\in N}
  \begin{tikzpicture}
     \pic[scale=0.8] (v) at (-0.6,0.15) {V1p=//n};
      \pic[scale=0.8] (w) at (0.6,-0.15) {W1p=/n/\tilde{m}};
       \node[irrep] at (-0.65,0.5) {$m\myinv{n}$};
      \node[irrep] at (1,0.2) {${n\tilde{m}}$};
    \node[irrep] at (-0.8,-0.6) {$\tilde{m}$};
    \draw[rounded corners,red] (v-mid)--++ (-0.15,0)--++(0,-0.4);
    \draw[rounded corners,red] (v-up)--++(0.1,0)--++(0,0.4);
  \draw[rounded corners,red] (w-down)--++(-0.1,0)--++(0,-0.2);
    \draw[rounded corners,red] (w-mid)--++ (0.15,0)--++(0,0.4);
     \draw[red] (v-down) to [out=0,in=180] (w-up);
    \node[tensor] (t) at (0,0) {};
    \draw (t)-++ (0,0.3)-++ (0,-0.3);
  \end{tikzpicture} \ ,
\end{equation*}
where now the gauge d.o.f.\ are supported in $\mathbb{C}[N]$. With these operators we construct the global projector $\P_N$ so that the partial gauging procedure is then $\G_N | \psi \rangle = \mathcal{P}_N \left(|\psi \rangle \ \bigotimes_{i}|v\rangle_{[i]_r} \right ) .$ We show here that the remaining global symmetry of the gauged state $\G_N |\psi \rangle$ is given by the quotient $G/N$.  The global symmetry operators now have to act on the matter and gauge d.o.f., i.e.\ they belong to $\mathcal{L}[\mathcal{H}^{\rm m} \otimes \mathcal{H}^{\rm g}]$, we propose the following:
\begin{equation}\label{eq:defUhat}
\hat{U}_g = \sum_{ \{ n_i \in N\} } \cdots
  \begin{tikzpicture}
\node[tensor] (t1) at (-0.8,0) {};
\draw[red] (t1)--++(-0.3,0);
\draw (-0.8,0.3)--(-0.8,-0.3);
 \pic[scale=0.8] (v) at (0.2,0.15) {V1p=//g};
      \pic[scale=0.8] (w) at (-0.2,-0.15) {W1p=/ g/};
      \node[] at (-0.15,-0.7) {${n_1}$};
      \node[] at (0.2,0.8) {${g n_1\myinv{g}}$};
\draw[red] (w-up) to [out=180,in=0] (t1);
\draw[red] (w-mid) to [out=0,in=180] (v-mid);
    \draw[rounded corners,red] (w-down)--++ (-0.1,0)--++(0,-0.3);
    \draw[rounded corners,red] (v-up)--++ (0.1,0)--++(0,0.3);
     \begin{scope}[shift={(1.6,0)}]
     \node[tensor] (t1) at (-0.8,0) {};
     \draw[red] (v-down) to [out=0,in=180] (t1);
\draw (-0.8,0.3)--(-0.8,-0.3);
  \pic[scale=0.8] (v) at (0.2,0.15) {V1p=//g};
      \pic[scale=0.8] (w) at (-0.2,-0.15) {W1p=/ g/};
      \node[] at (-0.15,-0.7) {${n_2}$};
      \node[] at (0.2,0.8) {${g n_2\myinv{g}}$};
\draw[red] (w-up) to [out=180,in=0] (t1);
\draw[red] (w-mid) to [out=0,in=180] (v-mid);
    \draw[rounded corners,red] (w-down)--++ (-0.1,0)--++(0,-0.3);
    \draw[rounded corners,red] (v-up)--++ (0.1,0)--++(0,0.3);
    \node[tensor] (t2) at (0.8,0) {};
    \draw[red] (v-down) to [out=0,in=180] (t2);
    \draw (0.8,0.3)--(0.8,-0.3);
    \draw[red] (t2)--++(0.3,0);
     \end{scope}
  \end{tikzpicture}
  \cdots
\end{equation}
We remind that since $N$ is normal in $G$, $gn\myinv{g}\in N$.
These operators satisfy the following relations (see proof in Appendix \ref{sec:proofs}):
\begin{equation}\label{eq:Uhatgrouprop}
    \hat{U}_g \hat{U}_h = \hat{U}_{gh}
\end{equation}
and
\begin{equation}\label{eq:UhatGNprop}
    \hat{U}_g \G_N = \G_N {U}_g, \ \forall g\in G
\end{equation}
and since $\G_N = \G_N {U}_n$ and  $\hat{u}^{[i]}_n\G_N = \G_N$ the global symmetry is given by $G/N$ because the whole symmetry is (compare with Eq.\ \eqref{eq:remLocSymmOnsite})
$$\hat{U}_g \bigotimes_{\{n_i\}} \hat{u}^{[i]}_{n_i} \  \G_N = \G_N {U}_{g'} \ ,
$$
for any $g' \in [g]$.

\subsection{On-site case and relation to previous gauging procedures}

In the following we show how our gauging procedure reduces to the one of Ref.\ \cite{Haegeman14} when considering on-site symmetries. For this one has to consider the global symmetry $U_g=\bigotimes_i u_g$ as an MPO with tensor
\begin{equation}\label{eq:tensoronsite}
 T_g = 
\begin{tikzpicture}
\draw (0,-0.4) -- (0,0.3);
\node[tensor] at (0,0) {};
\node[irrep] at (-0.2,0) {$u_g$};
\node[irrep] at (0.35,-0.6) {$|g\rangle \! \langle g|$};
\draw[red,preaction={draw, line width=2pt, white}] (-0.2,-0.2)--(0.6,-0.2);
\node[tensor] at (0.3,-0.2) {};
\end{tikzpicture} \ ,
\end{equation} 
where the virtual space is one dimensional and keeps track of the group element. The total virtual space of the MPO is then $\mathbb{C}[G]$ and the fusion tensors are

\begin{equation}\label{eq:fusiononsite}   W_{g,h} = 
\begin{tikzpicture}
\node[tensor,label=above:$|g \rangle$] (t1) at (0,0.3) {};
\draw[red] (t1) --++ (-0.35,0);
\node[tensor,label=below:$|h \rangle$] (t2) at (0,-0.3) {};
\draw[red] (t2) --++ (-0.35,0);
\node[tensor,label=above:$\langle gh |$] (t3) at (0.35,-0.05) {};
\draw[red] (t3) --++ (0.35,0);
\end{tikzpicture} \ ,
\end{equation} 
so that Eq.\ \eqref{eq:locsym}, the local symmetry operator,  is
$$ \hat{u}_g = \sum_{h,k} \ket{hg^{-1}}   \bra{h} \otimes u_g \otimes \ket{gk} \bra{k}  = R_g \otimes u_g \otimes L_g \ .  $$
The previous local symmetry operator together with the fact that the symmetrization is done by group averaging shows the equivalence to the procedure of Ref.\ \cite{Haegeman14}.

\subsection{Particularizing for MPS}

The previous gauging procedure has a very natural form when the MPO-symmetric state is an MPS $|\psi \rangle=|\psi_A \rangle$. We show here that applying the gauging procedure on a symmetric MPS gives rise to another MPS that can be written as follows:
\begin{equation}\label{eq:gaugedmps}
    \mathcal{G} | \psi_{A} \rangle = | \psi_{\tilde{A}B} \rangle \ .
\end{equation}
Using Eq.\ \eqref{eq:gaugeop}, the gauged MPS is written as
$$
\mathcal{G} | \psi_A \rangle =  \sum_{\{g_i\}}\ 
\begin{tikzpicture}
\node[] at (-0.6,0) {$\cdots$};
\node[] at (4.2,0) {$\cdots$};
\node[] at (-0.6,-0.5) {$\cdots$};
\node[] at (4.2,-0.5) {$\cdots$};
\draw[red] (-0.3,0) -- (0,0); 
\foreach \x in {0,1.3,2.6}{
\pic[scale=0.6] (v) at  ($ (\x,0) + (0.5,0) $) {V1p=//};
\draw[red] (v-mid) -- (\x,0); 
    \node[tensor] (t) at (\x,0) {};
    \node[tensor]  at (\x,-0.5) {};
    \draw (t)-++ (0,0.3)-++ (0,-0.5);
\draw[red,rounded corners] (v-up)--++(0.125,0)--++(0,0.25);
\draw[red] (v-down) to [out=0,in=180] ($(\x,0) + (1.3,0)$);
}
\draw  (-0.3,-0.5) -- (3.8,-0.5);
 \node[irrep] at (0.2,0) {$g_i$};
 \node[irrep] at (0.9,-0.1) {$g_{\scriptstyle i+1}$};
 \node[irrep] at (2.2,-0.1) {$g_{\scriptstyle i+2}$};
  \node[irrep] at (3.5,-0.1) {$g_{\scriptstyle i+3}$};
   \node[irrep] at (0.6,0.4) {$g_i\myinv{g}_{\scriptstyle i+1}$};
    \node[irrep] at (1.9,0.4) {$g_{\scriptstyle i+1}\myinv{g}_{\scriptstyle i+2}$};
    \node[irrep] at (3.3,0.4) {$g_{\scriptstyle i+2}\myinv{g}_{\scriptstyle i+3}$};
\end{tikzpicture} \ .
$$
To show Eq.\ \eqref{eq:gaugedmps} we first calculate the following 
\begin{align*}
\begin{tikzpicture}[baseline=-2mm]
\draw[red] (-0.3,0) -- (0,0); 
\draw[red] (1.3,0) -- (1.6,0); 
\draw (-0.3,-0.4) -- (1.6,-0.4);
\foreach \x in {0}{
\pic[scale=0.6] (v) at  ($ (\x,0) + (0.5,0) $) {V1p=//};
\draw[red] (v-mid) -- (\x,0); 
    \node[tensor] (t) at (\x,0) {};
    \draw (t)-++ (0,0.3)-++ (0,-0.4);
    \node[tensor] at (\x,-0.4) {};
\draw[red,rounded corners] (v-up)--++(0.125,0)--++(0,0.25);
\draw[red] (v-down) to [out=0,in=180] ($(\x,0) + (1.3,0)$);
}
\draw (1.3,0.3)--(1.3,-0.4);
\node[tensor] at (1.3,0) {};
\node[tensor] at (1.3,-0.4) {};
 \node[irrep] at (0.2,0) {$g_i$};
 \node[irrep] at (0.9,-0.1) {$g_{\scriptstyle i+1}$};
   \node[irrep] at (0.6,0.4) {$g_i\myinv{g}_{\scriptstyle i+1}$};
\end{tikzpicture}
 \ = \ \ &
\begin{tikzpicture} [baseline=-2mm]
\foreach \x in {0}{
\pic[scale=0.6] (v) at  ($ (\x,0) + (0.5,0) $) {V1p=//};
\draw[red,rounded corners] (v-up)--++(0.125,0)--++(0,0.25);
}
\pic[scale=0.8] (v1) at (0,-0.25)  {V2=//};
\pic[scale=0.8] (w1) at (1.3,-0.25)  {W2=//};
\draw (v1-down) -- (w1-down);
\draw[red] (v-down) to [out=0,in=180] (w1-up);
\pic[scale=0.8] (v1) at (1.9,-0.25)  {V2=/g_{\scriptstyle i+1}/};
\pic[scale=0.8] (w1) at (-0.6,-0.25)  {W2=/g_i/};
\node[tensor] at (-0.3,-0.25) {};
\node[tensor] at (1.6,-0.25) {};
\draw (-0.3,0)--(-0.3,-0.25);
\draw (1.6,0)--(1.6,-0.25);
 \node[irrep] at (0.2,0) {$g_i$};
 \node[irrep] at (0.9,-0.1) {$g_{\scriptstyle i+1}$};
   \node[irrep] at (0.6,0.4) {$g_i\myinv{g}_{\scriptstyle i+1}$};
\end{tikzpicture}
\\ =  \ \  & 
(L_{g_i\myinv{g}_{i+1}, g_{i+1}})^{-1} \ 
\begin{tikzpicture}[baseline=-1mm]
\pic[scale=0.8] (v1) at (2,0)  {V2=/g_{\scriptstyle i+1}/};
\pic[scale=0.8] (w1) at (-0.3,0)  {W2=/g_i/};
\foreach \x in {0}{
\pic[scale=0.6] (v) at  ($ (\x,0) + (0.7,0) $) {V2=//};
\draw (v-mid) -- (\x,0); 
    \node[tensor] (t) at (\x,0) {};
    \draw (t)-++ (0,0.3);
\draw[red,rounded corners] (v-up)--++(0.125,0)--++(0,0.25);
\draw (v-down) to [out=0,in=180] ($(\x,0) + (1.6,0)$);
}
\draw (1.7,0.3)--(1.7,0);
\node[tensor] at (1.7,0) {};
   \node[irrep] at (0.7,0.4) {$g_i\myinv{g}_{\scriptstyle i+1}$};
\end{tikzpicture}
\end{align*}
where in the first equality we have used Eq.\ \eqref{eq:actiontensors} and for the second Eq.\ \eqref{eq:Ldef}. Then, the tensor $B$ can be written as
\begin{equation}\label{Btensor}
B = \sum_{g,h} (L_{g\myinv{h}, h})^{-1}
  \begin{tikzpicture}
\draw[red,rounded corners] (-0.5,0)--++(0.2,0)--++(0,0.2);
     \pic[scale=0.8] (v) at (-0.5,-0.3) {V3=//};
     \node[irrep] at (-0.4,0.1) {$g\myinv{h}$};
     \node[irrep] at (-0.5,-1.2) {$|g\rangle \! \langle h |$};
     \draw (-1,-0.8)--(0,-0.8);
\node[tensor] at (-0.5,-0.8) {};
  \end{tikzpicture} ,
\end{equation}
and the new site tensor $\tilde{A} = A \otimes \id_G$ is just $A$ augmented with a trivial virtual space.

\subsection{Examples} 

In this section we exemplify our gauging procedure using the MPOs coming from the symmetries of boundary theories of 2D topological orders. In particular, let us consider 2D PEPS describing topological orders based on groups \cite{Bultinck17A} whose main property is the virtual invariance of the PEPS tensors under MPO representations of a finite group. The MPOs that we are interested in are anomaly free, i.e. they are characterized by a trivial 3-cocycle, and the corresponding PEPS describe quantum double models. So if we consider a 1D state living at the virtual boundary of the PEPS, this will inherit the MPO virtual symmetries of the PEPS as global symmetries. In what follows we use the MPO group representations appearing in  Ref.\ \cite{Garre22} which are based on the ones of Ref.\ \cite{Bultinck17A}.

The local Hilbert space of these MPOs is $\mathbb{C}[G]\otimes \mathbb{C}[G]$. Since the sites have an internal tensor product structure we label their left and right components as $[i]_l$ and $[i]_r$ respectively. Let $O_g$ denotes the PBC MPO described by the MPO tensor $T_g$:

\begin{equation}\label{tensorexam}
T_g = 
\begin{tikzpicture}
\draw (-0.2,-0.5)--(-0.2,0.75);
\draw (0.2,-0.5)--(0.2,0.75);
  \node[tensor,label=right:$L_g$]  at (0.2,0.5) {};
   \node[tensor,label=left:$L_g$]  at (-0.2,0.5) {};
\draw[red] (-0.2,-0.2)--(-0.6,-0.2);
\draw[red] (0.2,-0.2)--(0.6,-0.2);
\end{tikzpicture} \ , 
\end{equation}
where the virtual space is $\mathbb{C}[G]$, for every $g$, and every three-line intersection corresponds to a $|G|$-dimensional delta tensor. In particular, the PBC MPO corresponding to the trivial element $e$ is the projector
$$ 
O_e = \bigotimes_i \Pi^{[i]_r,[i+1]_l} = \ 
\cdots \
\begin{tikzpicture}[scale=0.8]
\foreach \x in {-1,0,1}{
\draw[densely dotted,rounded corners] (\x+0.2,-0.35) rectangle (\x+0.8,0.35);
}
\node at (0.5, 0.45) {$[i]$};
\node at (0.25, -0.5) {$[i]_l$};
\node at (0.85, -0.5) {$[i]_r$};
\foreach \x in {-1,0,1,2}{
\begin{scope}[shift={(\x,0)}]
\draw (-0.3,-0.25)--(-0.3,0.25);
\draw[red] (-0.3,0) -- (0.3,0);
\draw (0.3,-0.25)--(0.3,0.25);
\end{scope}}
\end{tikzpicture}
\ \cdots,
$$
where $\Pi= \sum_g |g,g\rangle\langle g,g|$. 
We note that the projector $\Pi$ acts on two consecutive sites, $i$ and $i+1$. For any $g \in G$ we can write
$$ O_g = \left ( \bigotimes_i L_g^{[i]_l}\otimes L_g^{[i]_r} \right ) O_e \ .$$

These operators satisfy $O_g O_h= O_{gh}$ and $O_g^\dagger =O_{g^{-1}}$, so they form a representation of $G$ on the subspace $O_e \cdot (\mathbb{C}[G]\otimes \mathbb{C}[G])^{\otimes n } $ of the full Hilbert space $(\mathbb{C}[G]\otimes \mathbb{C}[G]) ^{\otimes n }$.

The fusion tensors are given by 
\begin{equation}\label{ftexam}
\begin{tikzpicture}
\draw[red] (-0.65, 0) -- (0, 0);
\node at (-0.4, 0.15) {$gh$};
\node at (-0.95, 0) {$\langle k'|$};
\draw[red] (0.6, 0.35) -- (0, 0.35);
\node at (0.4, 0.5) {$g$};
\node at (0.85, 0.35) {$|q\rangle$};
\draw[red] (0.6, -0.35) -- (0, -0.35);
\node at (0.4, -0.2) {$h $};
\node at (0.85, -0.35) {$|k \rangle $};
\draw[rounded corners, fill= black] (-0.15,-0.65) rectangle (0.15,0.65);
\end{tikzpicture}
= \delta_{q,hk} \cdot \delta_{k,k'}  \ .
\end{equation}
It is easy to verify that these fusion tensors satisfy Eq.\ \eqref{eq:fusiontensors} and Eq.\ \eqref{3cocytriv}. Using these fusion tensors and the MPO tensor of Eq.\ \eqref{tensorexam} we obtain that the local symmetry operators defined in  Eq.\ \eqref{eq:locsym} are 
$$ \hat{u}^{[i]}_g  =
\begin{tikzpicture}
\draw[red] (-1.2, 0.25) -- (-1.2, -0.5);
\node[tensor,label={[shift={(-0.2,-0.05)}] $R_g$}] at (-1.2,0) {};
\draw[red] (-0.8, 0.25) -- (-0.8, -0.5);
\node[tensor,label={[shift={(-0.2,-0.05)}] $L_g$}] at (-0.8,0) {};
\draw[red] (-0.8, -0.35) -- (-0.25, -0.35);
\draw[] (-0.25, 0.25) -- (-0.25, -0.5);
\draw[densely dotted,rounded corners] (-0.4,-0.55) rectangle (0.4,0.4);
\node at (0, 0.55) {$[i]$};
%
\draw[densely dotted,red, rounded corners] (-0.65,-0.55) rectangle (-1.35,0.4);
\node at (-1, 0.55) {$\{i-1\}$};
\node[scale=0.7] at (-1.4, -0.7) {$\{i-1\}_l$};
\node[scale=0.7] at (-0.7, -0.7) {$\{i-1\}_r$};
\draw[densely dotted,red, rounded corners] (0.65,-0.55) rectangle (1.35,0.4);
\node at (1, 0.55) {$\{i\}$};
\node at (1.3, -0.7) {$\{i\}_l$};
\node at (0.8, -0.7) {$\{i\}_r$};
\node[tensor,label={[shift={(-0.2,-0.05)}] $L_g$}] at (-0.25,0) {};
\draw[] (0.25, 0.25) -- (0.25, -0.5);
\node[tensor,label={[shift={(-0.2,-0.05)}] $L_g$}] at (0.25,0) {};
\draw[red] (0.8, -0.35) -- (0.25, -0.35);
\draw[red] (0.8, 0.25) -- (0.8, -0.5);
\node[tensor,label={[shift={(-0.2,-0.05)}] $L_g$}] at (0.8,0) {};
\draw[red] (1.2, 0.25) -- (1.2, -0.5);
\end{tikzpicture} \  \ \forall g\in G \ ,
$$
where the new gauge d.o.f. are depicted in red and its Hilbert space is $\mathbb{C}[G]\otimes \mathbb{C}[G]$.
The new gauge d.o.f.\ have also a tensor product structure, so we denote their component as $\{i\}_l$ and $\{i\}_r$ (the gauge site $\{i\}$ is located between the matter sites $[i]$ and $[i+1]$). 

Using these local operators we construct the gauging map defined in Eq.\ \eqref{eq:gaugeop}:
$$ \G = \sum_{\{g_i\}} \cdots
\begin{tikzpicture}
\draw[] (-1.6, 0.25) -- (-1.6, -0.5);
\node[tensor] at (-1.6,0) {};
\node[scale=0.75] at (-1.8,0.175) {$L_{g_1}$};
\draw[red] (-1.2, 0.25) -- (-1.2,0);
\node[tensor] at (-1.2,0) {};
\node[scale=0.65] at (-1.2,-0.2) {$|g_1\myinv{g}_2\rangle$};
\draw[red] (-0.8, 0.25) -- (-0.8, -0.35);
\node[tensor] at (-0.8,0) {};
\node[scale=0.75] at (-1,0.175) {$L_{g_2}$};

\draw[red] (-1.6, -0.35) -- (-0.25, -0.35);
\draw[] (-0.25, 0.25) -- (-0.25, -0.5);
\node[tensor] at (-0.25,0) {};
\node[scale=0.75] at (-0.45,0.175) {$L_{g_2}$};

\draw[] (0.25, 0.25) -- (0.25, -0.5);
\node[tensor] at (0.25,0) {};
\node[scale=0.75] at (0.05,0.175) {$L_{g_2}$};

\draw[red] (1.6, -0.35) -- (0.25, -0.35);
\draw[red] (0.8, 0.25) -- (0.8, -0.35);
\node[tensor] at (0.8,0) {};
\node[scale=0.75] at (0.55,0.175) {$L_{g_2}$};
\draw[red] (1.2, 0.25) -- (1.2, 0);
\node[tensor] at (1.2,0) {};
\node[scale=0.65] at (1.2,-0.2) {$|g_2\myinv{g}_3\rangle$};
\draw[] (1.6, 0.25) -- (1.6, -0.5);
\node[tensor] at (1.6,0) {};
\node[scale=0.75] at (1.4,0.175) {$L_{g_3}$};
\end{tikzpicture} \ \cdots $$ 
In order to better understand this example we compare it with the gauging of on-site symmetries. In that case, when the local operators of the global symmetry are $u_g = L_g$, the gauging map has the form
$$ \G_{ {\rm on}} = \sum_{\{g_i\}} \cdots
\begin{tikzpicture}
\draw [] (-1, -0.35) -- (-1, 0.35);
  \node[tensor,label=left:$L_{g_1}$]  at (-1,0) {};
  \draw [red] (0, 0) -- (0, 0.35);
  \node[tensor,label=below:$| g_1 \myinv{g}_2 \rangle$]  at (0,0) {};
  \draw [] (1, -0.35) -- (1, 0.35);
  \node[tensor,label=left:$L_{g_2}$]  at (1,0) {};
  \draw [red] (2, 0) -- (2, 0.35);
  \node[tensor,label=below:$| g_2 \myinv{g}_3 \rangle$]  at (2,0) {};
  \draw [] (3, -0.35) -- (3, 0.35);
  \node[tensor,label=left:$L_{g_3}$]  at (3,0) {};
\end{tikzpicture} \ \cdots $$ 
If we then consider the state $|+\rangle_G^{\otimes n }= (\frac{1}{|G|}\sum_g |g\rangle) ^{\otimes n }$ invariant under $L_g^{\otimes n}$ and we gauge it we obtain:
$$ \G_{\rm on} |+\rangle_G^{\otimes n} = \left (|+\rangle_G^{\otimes n} \right )_\mathfrak{m}\otimes \left (\bigotimes_{g_i} |g_i \myinv{g}_{i+1} \rangle \right )_\mathfrak{g} \ ,$$
where we can see that matter and gauge fields are decoupled. It is interesting to note that the resulting gauged state has an emergent global symmetry acting only on the gauge fields: $\prod_i Z_\alpha^{\{i\}}$, where the single-site operator $Z_\alpha$ transforms according to a one-dimensional irrep of $G$: $Z_\alpha |g\rangle  = \alpha(g) |g\rangle$ where $\alpha(g) \alpha(h) = \alpha(gh)$. If the group $G$ is Abelian their irreps are one-dimensional and form a group isomorphic to $G$ so they satisfy $\alpha\cdot \beta = \alpha \beta \in Irrep(G)$ so that $Z_\alpha Z_\beta = Z_{\alpha \beta}$. Then, the new emergent symmetry is a representation of $G$. The case for non-Abelian groups turns out to be more complex where the resulting symmetry is a representation of the fusion category Rep($G$) and it is outside the scope of this work. Emergent symmetries have recently gained attention due to their connection with dualities, see Refs.\ \cite{Bhardwaj18,lootens21dualA} and references therein.

We can now compare this with the situation for our gauging map of non-on-site symmetries. To do so, we first consider an MPO-symmetric state $|W\rangle = \prod_i |\omega\rangle_{[i]_r,[i+1]_l}$, where
$|\omega\rangle = \frac{1}{|G|}\sum_g |g,g\rangle$ so that $O_g |W\rangle = |W\rangle $ for all $g\in G$. Then, after gauging we obtain
$$ \G |W\rangle = \sum_{\{g_i\}} \cdots
\begin{tikzpicture}
\draw[] (-1.6, 0.25) -- (-1.6, -0.35);
\node[tensor] at (-1.6,0) {};
\node[scale=0.75] at (-1.9,-0.175) {$L_{g_1\myinv{g}_2}$};
\draw[red] (-1.2, 0.25) -- (-1.2,0);
\node[tensor] at (-1.2,0) {};
\node[scale=0.65] at (-1.2,-0.2) {$|g_1\myinv{g}_2\rangle$};
\node[tensor] at (-0.95,-0.55) {};
\node[] at (-0.95,-0.75) {$| \omega \rangle$};
\draw[red,rounded corners] (-0.8, 0.25) -- (-0.8, -0.25)-- (-0.25, -0.25);
\draw[rounded corners] (-1.6, 0.25)-- (-1.6, -0.55)-- (-0.25, -0.55)-- (-0.25, 0.25);
\begin{scope}[shift={(2,0)}]
\draw[] (-1.6, 0.25) -- (-1.6, -0.35);
\node[tensor] at (-1.6,0) {};
\node[scale=0.75] at (-1.9,-0.175) {$L_{g_2\myinv{g}_3}$};
\draw[red] (-1.2, 0.25) -- (-1.2,0);
\node[tensor] at (-1.2,0) {};
\node[scale=0.65] at (-1.2,-0.2) {$|g_2\myinv{g}_3\rangle$};
\node[tensor] at (-0.95,-0.55) {};
\node[] at (-0.95,-0.75) {$| \omega \rangle$};
\draw[red,rounded corners] (-0.8, 0.25) -- (-0.8, -0.25)-- (-0.25, -0.25);
\draw[rounded corners] (-1.6, 0.25)-- (-1.6, -0.55)-- (-0.25, -0.55)-- (-0.25, 0.25);
\end{scope}
\end{tikzpicture}
\ \cdots
$$
In contrast with the on-site case, the resulting gauged state is not a tensor product state between gauge and matter fields, i.e. they cannot be decoupled. This fact cannot be obtained from the general theory that we have derived above and it has to be checked case-by-case. Apart from  this difference we notice that, similarly to the on-site case,  there is  an emerging global symmetry acting on the gauge fields; $\prod_i Z_\alpha^{\{i\}_l}( \G |W \rangle) =  \G | W\rangle $, where $\alpha$ is any one-dimensional irrep (so, as before, for Abelian groups the new global symmetry is a representation of $G$).

\section{Localizing anomalous MPOs}\label{sec:MPOtopo}

An anomalous MPO representation is characterized by fusion tensors that satisfy Eq.\ \eqref{eq:3cocycle} with $\omega$ being non-trivial so that Eq.\ \eqref{3cocytriv} and Eq.\ \eqref{eq:trivial3ortho} are not satisfied for all group elements.

If we define the local operators as in Eq.\ \eqref{eq:locsym} we find out that they no longer satisfy the group product relations because of the non-trivial $\omega$ factors in Eq.\ \eqref{eq:3cocycle}. Alternatively, we could try to  use the following condition from associativity of the product: 
\begin{equation}\label{rawrels}
  \begin{tikzpicture}
    \pic[scale=0.9] (w1) at (-1,0) {V1=/gh/k};
    \node[irrep] at (1.3,0) {${ghk}$};
    \node[irrep] at (-1.3,0) {${ghk}$};
    \pic[scale=0.9] (w2) at (w1-up) {V1=/g/h};
    \pic[scale=0.9] (v1) at (1,0) {W1=/gh/k};
    \pic[scale=0.9] (v2) at (v1-up) {W1=/g/h};
  \end{tikzpicture} =
  \begin{tikzpicture}
    \pic[scale=0.9] (w1) at (-1,0) {V1=/g/hk};
    \pic[scale=0.9] (w2) at (w1-down) {V1=/h/k};
    \node[irrep] at (1.3,0) {${ghk}$};
    \node[irrep] at (-1.3,0) {${ghk}$};
    \pic[scale=0.9] (v1) at (1,0) {W1=/g/hk};
    \pic[scale=0.9] (v2) at (v1-down) {W1=/h/k};
  \end{tikzpicture} 
\end{equation}
to define the local operators as follows (correlating the left and the right side):
\begin{equation}\label{symop}
\check{u}_g =\sum_h
  \begin{tikzpicture}
     \pic[scale=0.8] (v) at (-0.6,0) {V1p=/g/h};
      \pic[scale=0.8] (w) at (0.6,0) {W1p=/g/h};
      \node[irrep] at (0.9,0.45) {${gh}$};
    \node[irrep] at (-0.9,0.45) {${gh}$};
    \draw[rounded corners,red] (v-down)--++ (0.15,0)--++(0,-0.2);
    \draw[rounded corners,red] (v-mid)--++(-0.15,0)--++(0,0.5);
  \draw[rounded corners,red] (w-down)--++(-0.15,0)--++(0,-0.2);
    \draw[rounded corners,red] (w-mid)--++ (0.15,0)--++(0,0.5);
\draw[red] (v-up)--(w-up);
    \node[tensor] (t) at (0,0.21) {};
    \draw (t)-++ (0,0.3)-++ (0,-0.3);
  \end{tikzpicture} 
\end{equation}
which satisfy $\check{u}_g \check{u}_h = \check{u}_{gh}$.

The main problem to proceed from here is that if we add the new d.o.f.\ as in the previous section, $[i]_r\equiv[i+1]_l$, we cannot ensure that $\check{u}^{[i]}_g$ and $\check{u}^{[i+1]}_h$ commute. 

In order to make them commute we could split the gauge Hilbert space: the left and right local operators have to be supported in different Hilbert spaces, that is, we introduce $\H_{[i]_r} \otimes \H_{ [i+1]_l}$ between matter sites $i$ and $i+1$. 
Since the operators $\check{u}^{[i]}_{g}$ only acts on the sites $[i]_r\otimes i\otimes [i]_l$ and do not interact with any other site, the new d.o.f.\ can be considered as matter fields. We notice that whole symmetry is realized in a tensor product form:
$$\bigotimes_i \check{u}^{[i]}_{g_i} \ . $$

However in the next section we show how we can use $\check{u}_{g}$ to symmetrize states and operators satisfying similar relations as the previous section.

\subsection{ Symmetrizing (group averaging) the localized anomaly }

We introduce gauge fields by decorating each site of the chain with a left and a right Hilbert space with $\mathcal{H}_{[i]_r}= \mathcal{H}_{[i]_l}= \mathbb{C}^\chi$. But now the total Hilbert space of the gauge fields is $\mathcal{H}^{\rm g} = \bigotimes_{i}(\mathcal{H}_{[i]_r}\otimes \mathcal{H}_{[i+1]_l})$.

We define the local operators acting on $\hat{\mathcal{H}}_i^{\rm local} = \mathcal{H}_{[i]_l} \otimes \mathcal{H}_i \otimes \mathcal{H}_{[i]_r} $ as in Eq.\ \eqref{symop}.
In general, $\check{u}_e$ is not the identity on $\tilde{\mathcal{H}}_{\rm local}$ but it is a projector $\check{u}^2_e = \check{u}_e$. Then, $\check{u}_g$ is a representation on the subspace $\check{u}_e \cdot \hat{\mathcal{H}}^{\rm local}$. 

The local projectors are defined using $\check{u}_g$ by group averaging so that the symmetrization is given by

$$\mathcal{G}_\phi | \psi \rangle = \mathcal{P} \left(|\psi \rangle \otimes | \phi \rangle  \right ),$$
where $| \phi \rangle$ is any state in $\mathcal{H}^{\rm g}$. It can be checked that for $| V\rangle =\bigotimes_{i}|v\rangle_{[i]_r} |v\rangle_{[i+1]_l}$ we have
\begin{equation*}
\mathcal{G}_V | \psi \rangle =  \sum_{g_1,g_2,\cdots} \frac{1}{|G|}\left (T_{g_i}^{[i]} \right)_{\alpha,\beta} |\psi \rangle |\alpha\rangle_ {[i]_l} |\beta \rangle_{[i]_r} \ ,    
\end{equation*}
or in graphical notation:
\begin{equation}\label{trivgau}
\mathcal{G}_V | \psi \rangle = \prod_i \sum_{\{g_i\}}\ 
  \begin{tikzpicture}
  \draw[rounded corners]  (-1.5,-0.3) rectangle (1.5,0);
   \draw[rounded corners,red] (-0.85,0.6)--++(0,-0.3)--++(0.7,0)--++(0,0.3);
      \draw[rounded corners,red] (0.15,0.6)--++(0,-0.3)--++(0.7,0)--++(0,0.3);
  \node[tensor]  at (-0.5,0.3) {};
    \draw[] (-0.5,0)--(-0.5,0.6);
    \node[tensor]  at (0.5,0.3) {};
    \draw[] (0.5,0)--(0.5,0.6);
    \node[irrep] at (-1.2,0.1) {$\cdots$};
    \node[irrep] at (1.2,0.1) {$\cdots$};
     \node[irrep] at (-0.75,0.035) {$g_i$};
      \node[irrep] at (0.2,0.035) {$g_{\scriptstyle i+1}$};
        \node[] at (0,-0.15) {$ \psi $};
      \end{tikzpicture}   \ .
\end{equation}

For convenience, we define the state $|\Omega \rangle = \bigotimes_{i} |\omega \rangle_{[i]_r,[i+1]_l} \in \mathcal{H}^{\rm g}$, where $|\omega \rangle = \frac{1}{\sqrt{\chi}} \bigoplus_g  |\omega_g \rangle$ and $|\omega_g \rangle = \sum_{s=1}^{\chi_g}|s,s\rangle $. Then, it can be seen that 
\begin{equation*}
  \langle \Omega|\mathcal{G}_{\Omega} U_g | \Psi \rangle = \langle \Omega|\mathcal{G}_{\Omega} | \Psi \rangle \ ,  
\end{equation*}
 for any $| \Psi \rangle \in  \mathcal{H}^{\rm m}$. This arises from the fact that $\langle \Omega| \mathcal{P}  | \Omega \rangle \propto \sum_g U_g$.

For operators, we define the map $\Gamma[O] = S \circ \mathcal{E}_\Lambda[O] $, where
the enlarging map is
$$\mathcal{E}_\Lambda[O] =  O  \bigotimes_{i\in \Lambda}( |v\rangle_{[i]_l} \langle v|_{[i]_l} \otimes |v\rangle_{[i]_r} \langle v|_{[i]_r}) \ , $$ 
and $S$ is defined using the $\check{u}_g$ operators. Interestingly, the symmetrizing procedure of operators and states is compatible in the following sense:
\begin{equation*}
\langle \Omega| \Gamma[O] \mathcal{G}_{\Omega} | \psi \rangle = \langle \Omega|\mathcal{G}_{\Omega} (O | \psi \rangle) \ ,
\end{equation*}
where we have assumed that the state $| \psi \rangle$ and the operator $O$ commutes with the MPO $U_g$ for all $g\in G$. 

\subsection{On-site case and relation to previous gaugings}

In the following we show how the previous symmetrization map for on-site symmetries is connected to the gauging procedure of Ref.\cite{Haegeman14} by performing a renormalization procedure.

We first consider the total Hilbert space of the gauge fields, $\bigotimes_{i}\mathcal{H}_{[i]_r}\otimes \mathcal{H}_{[i+1]_l}$, and we map isometrically the even sites $\mathcal{H}_{[2i]_r}\otimes \mathcal{H}_{[2i+1]_l}$ to $\mathcal{H}_{[i]}$ and project out the odd sites. Then, we compress isometrically the matter fields between the projected out gauge fields: $\mathcal{H}_{2i-1}\otimes \mathcal{H}_{2i}\to \mathcal{H}_{i}$, so that the total Hilbert space of the gauge and matter fields is $\bigotimes_{i}\mathcal{H}_{i}\otimes \mathcal{H}_{[i]}$.

The MPO representation of on-site symmetries is given in Eq.\ \eqref{eq:tensoronsite} and Eq.\ \eqref{eq:fusiononsite} so that the local operators of Eq.\ \eqref{symop} for this case have the form
$$ \check{u}_g =\sum_h  | gh \rangle \langle h| \otimes u_g \otimes   | gh \rangle \langle h|. $$
The previously described transformation on the Hilbert space maps $\check{u}_g\otimes \check{u}_g$ to $ L_g \otimes u_g \otimes L_g$, where $L_g$ is the left regular representation (isomophic to the right regular representation). 

\subsection{Particularizing for MPS}

The previous gauging procedure has a very natural form when the MPO symmetric state is an MPS $|\psi \rangle=|\psi_A \rangle$. In this section we show that the resulting gauged state of Eq.\ \eqref{trivgau} is also an MPS with the following form: 
$$ \mathcal{G}_V | \psi_A \rangle = \sum_\alpha|\psi_{AB_\alpha} \rangle \ ,$$
where $\alpha$ stands for the blocks of $A=\bigoplus_\alpha A_\alpha$ and the tensor $B_\alpha$, whose explicit form will be given, corresponds to the gauge fields. It is important to note that in this case the MPOs perform a permutation of the injective blocks $U_g | \psi_{A_x} \rangle = | \psi_{A_y} \rangle$ for some $y$ that we denote by $y\equiv g{\cdot}x$.

Let us now show the explicit form of $\mathcal{G}_V | \psi_A \rangle = \sum_\alpha\mathcal{G}_V | \psi_{A_\alpha} \rangle$, to do so we calculate $\mathcal{G}_V | \psi_{A_x} \rangle$ as

\begin{align*}
\mathcal{G}_{V} | \psi_{A_x} \rangle & = \prod_i \sum_{\{g_i\}} \cdots 
  \begin{tikzpicture}
  \draw  (-1,-0.2) -- (1,-0.2);
   \draw[rounded corners,red] (-0.85,0.6)--++(0,-0.3)--++(0.7,0)--++(0,0.3);
      \draw[rounded corners,red] (0.15,0.6)--++(0,-0.3)--++(0.7,0)--++(0,0.3);
  \node[tensor]  at (-0.5,0.3) {};
  \node[tensor,label=below:$A_{x}$]  at (-0.5,-0.2) {};
    \draw[] (-0.5,-0.2)--(-0.5,0.6);
    \node[tensor]  at (0.5,0.3) {};
    \node[tensor,label=below:$A_{x}$]  at (0.5,-0.2) {};
    \draw[] (0.5,-0.2)--(0.5,0.6);
     \node[] at (-0.8,0.12) {$g_i$};
      \node[] at (0.15,0.12) {$g_{\scriptstyle i+1}$};
      \end{tikzpicture}  
      \cdots \\
      & =  \prod_i \sum_{\{g_i\}} \cdots
         \begin{tikzpicture}
  \draw[red,rounded corners] (-0.5,-0.10)--++(0.4,0)--++(0,0.4);
\draw[red,rounded corners] (0.5,-0.10)--++(-0.4,0)--++(0,0.4);
     \pic[scale=0.6] (v) at (-0.45,-0.3) {V3=//};
      \pic[scale=0.6] (w) at (0.45,-0.3) {W3=//};
\draw (v-down)--(w-down);
\node[irrep] at (0,-0.7) {$x$};
\node[] at (0.3,0.4) {$g_{i+1}$};
\node[] at (-0.35,0.4) {$g_{i}$};
\node[tensor] (t) at (1,-0.30) {};
\draw (t) --++ (0,0.3);
\draw (t) --++ (0.30,0)--++(-0.6,0);
\node[tensor,label=below:$A_{g_{i}{\cdot}x}$] (t) at (-1,-0.30) {};
\node[] at (1.2,-0.60) {$A_{g_{i+1}{\cdot}x}$};
\draw (t) --++ (0,0.3);
\draw (t) --++ (-0.30,0)--++(0.6,0);
\node[] at (1.6,-0.1) {$\cdots$};
      \end{tikzpicture} 
      \\
      & \equiv |\psi_{AB_x} \rangle \ ,
\end{align*}
where in the second equality we have used 

\begin{equation*}
  \begin{tikzpicture}
    \draw[red] (-0.5,0)--(0.5,0);
     \node[irrep] at (0.55,0) {$g$};
        \node[irrep] at (0.55,-0.5) {$x$};
    \draw (-0.5,-0.5)--(0.5,-0.5);
    \node[tensor] (t) at (0,0) {};
    \node[tensor] (t) at (0,-0.5) {};
    \draw (0,-0.5) -- (0,0.3);
  \end{tikzpicture} 
\ =  \ 
  \begin{tikzpicture}
    \pic[] (v) at (0.5,0) {V2=y/g/x};
    \pic[] (w) at (-0.5,0) {W2=y/g/x};
    \node[tensor] (t) at (0,0) {};
    \draw (0,0) -- (0,0.3);
  \end{tikzpicture} \ ,
\end{equation*} 
so that the tensor $B_x$ has the following form:
\begin{equation*}
B_x = \sum_{g,h}
  \begin{tikzpicture}
\draw[red,rounded corners] (-0.5,0)--++(0.4,0)--++(0,0.4);
\draw[red,rounded corners] (0.5,0)--++(-0.4,0)--++(0,0.4);
     \pic (v) at (-0.5,-0.3) {V3=g{\cdot}x/g/};
      \pic (w) at (0.5,-0.3) {W3=h{\cdot}x/h/};
\draw (v-down)--(w-down);
\node[irrep] at (0,-0.6) {$x$};
  \end{tikzpicture} .
\end{equation*}
It is instructive to rewrite $|\psi_{AB_x} \rangle = |\psi_{C_x} \rangle$ where
$$
C_x = \sum_{g_i}
  \begin{tikzpicture}
  \draw[red,rounded corners] (1.7,0)--++(0.4,0)--++(0,0.4);
\draw[red,rounded corners] (0.5,0)--++(-0.4,0)--++(0,0.4);
     \pic (v) at (1.7,-0.3) {V3=//x};
      \pic (w) at (0.45,-0.3) {W3=//x};
\node[] at (0.3,0.4) {$g_{i}$};
\node[] at (1.9,0.4) {$g_{i}$};
\node[tensor,label=below:$A_{g_i{\cdot}x}$] (t) at (1.1,-0.30) {};
\draw (t) --++ (0,0.3);
\end{tikzpicture}
\ ,
$$
which shows directly the triviality of the state and its invariance under the action of $\check{u}_g^{[i]}$.

\section{Conclusions and outlook}\label{outlook}
In this work we have shown explicitly how to  gauge non-anomalous symmetries at the level of quantum states in 1D. The main assumption is that the symmetry is realized by an MPO, whose local tensor structure allows for the construction of local symmetry operators with the same group relations as the MPOs, see Fig.\  \ref{fig:sketch}.\ for a sketch. 
Those local operators allow us to construct a projector to the gauge invariant subspace by a group averaging procedure. While showing that the resulting gauging procedure satisfies all the desired properties as in the on-site symmetry case, we repeatedly relied on the trivial 3-cocycle relation Eq.\ \ \eqref{3cocytriv}---i.e.\ the absence of an anomaly.
This sheds light on the question of  what is it that  obstructs the gauging of anomalous symmetries. For anomalous MPO symmetries a non-trivial phase appears in all the required relations. We explored the possibility of fixing this obstruction by localizing the operators which resulted in trivial (tensor product) symmetries.

Some open questions arise from this work. It would be interesting to see how MPO symmetries can be gauged at the Hamiltonian level and if our procedure can be generalized to non-anomalous symmetries beyond the group case, e.g. semi-simple Hopf algebras. We left these questions for future work.

In Ref.\cite{Bhardwaj18}, the authors study what is the structure of the remaining global symmetry after gauging a non-anomalous subgroup. We are able to do it just in the case where the whole symmetry is non-anomalous, obtaining results similar to those of as the on-site case. 
In Ref.\cite{Ji21}, the authors propose a gauging procedure for anomalous symmetries in a boundary by considering the boundary theory of the gauged bulk. We wonder if our symmetrization for anomalous symmetries matches with their proposal and if this implies a relation between gauging or symmetrization procedures and the bulk-boundary correspondence.

Our procedure of constructing local operators from MPOs is based on the existence of fusion tensors that decompose the product of local MPO tensors. This could be generalized to 2D (non-anomalous) operators with a tensor network structure, the so-called projected entangled pair operators \cite{reviewPEPS}, satisfying some group relations using their generalized fusion tensors.

\subsection*{Acknowledgements}
The authors deeply thank Erez Zohar for encouraging  and enlightening discussions. JGR thanks Laurens Lootens for useful comments on the manuscript.
This work has been partially supported by the European Research Council (ERC) under the European Union’s Horizon 2020 research and innovation programme through the ERC-CoG SEQUAM (Grant Agreement No.\ 863476).
\appendix

\section{Some proofs}\label{sec:proofs}

In this section we prove Eqs.\ \eqref{compgau},  \eqref{eq:Uhatgrouprop} and \ \eqref{eq:UhatGNprop}. We will show Eq.\ \eqref{compgau} for a three-site operator whose gauging procedure looks as follows:
$$
 \Gamma[O] =  \sum_{g_1,g_2,g_3}
\begin{tikzpicture}
\draw[red,rounded corners]  (0,0)--++(-0.3,0) --++ (0,0.25); 
\draw[red,rounded corners] (2.6,0) --++ (0.3,0)--++(0,0.2); 
\foreach \x in {0,1.3}{
\pic[scale=0.6] (v) at  ($ (\x,0) + (0.5,0) $) {V1p=//};
\draw[red] (v-mid) -- (\x,0); 
    \node[tensor] (t) at (\x,0) {};
    \draw (t)-++ (0,0.3)-++ (0,-1.2);
\draw[red,rounded corners] (v-up)--++(0.125,0)--++(0,0.25);
\draw[red] (v-down) to [out=0,in=180] ($(\x,0) + (1.3,0)$);
}
\draw (2.6,0.3)--(2.6,-1.2);
\node[tensor] at (2.6,0) {};
 \node[irrep] at (0.2,0) {$g_1$};
 \node[irrep] at (0.9,-0.1) {$g_2$};
 \node[irrep] at (2.2,-0.1) {$g_3$};
   \node[irrep] at (0.6,0.4) {$g_1\myinv{g}_{2}$};
    \node[irrep] at (1.9,0.4) {$g_{2}\myinv{g}_{3}$};
\draw[rounded corners, fill=white]  (-0.5,-0.35) rectangle (2.9,-0.65);
\node[] at (1.1,-0.5) {$ O $};
\draw[red,rounded corners]  (0,-1)--++(-0.3,0) --++ (0,-0.2); 
\draw[red,rounded corners] (2.6,-1) --++ (0.3,0)--++(0,-0.25);
\foreach \x in {0,1.3}{
\pic[scale=0.6] (w) at  ($ (\x,-1) + (0.5,0) $) {W1p=//};
\draw[red] (w-up) to [out=180,in=0] (\x,-1); 
    \node[tensor] (t) at (\x,-1) {};
\draw[red,rounded corners] (w-down)--++(-0.125,0)--++(0,-0.25);
\draw[red] (w-mid) -- ($(\x,-1) + (1.3,0)$);
}
\node[tensor] at (2.6,-1) {};
 \node[irrep] at (-0.2,-1.0) {$\myinv{g}_1$};
 \node[irrep] at (0.9,-1.05) {$\myinv{g}_2$};
 \node[irrep] at (2.2,-1.05) {$\myinv{g}_3$};
   \node[irrep] at (0.5,-1.7) {$g_1\myinv{g}_{2}$};
    \node[irrep] at (1.6,-1.7) {$g_{2}\myinv{g}_{3}$};
    \draw[fill=black] (-0.23,0) rectangle (-0.23-0.1,0+0.1);
    \node[] at (-0.25,-0.2) {$Z_{g_1}$};
    \draw[fill=black] (2.83,-1.1) rectangle (2.83+0.1,-1.1+0.1);
    \node[] at (2.9,-0.85) {$Z_{g_3}$};
\end{tikzpicture} \ ,
$$
where the $Z$ matrices come from the use of Eq.\ \eqref{eq:Zmatrixdef}.

Let us calculate directly $\Gamma[O] \mathcal{G} | \psi \rangle$ for a three-site operator $O$:
\begin{equation}\label{eq:GOGpsi}
  \sum_{\substack{g_1,g_2,g_3, \{k_i\} \\  \myinv{g}_{1}= k_0\myinv{k}_{1} \\ g_1\myinv{g}_{2}= k_1\myinv{k}_{2} \\ g_2\myinv{g}_{3}= k_2\myinv{k}_{3} \\ {g}_{3}= k_3\myinv{k}_{4}}}
\begin{tikzpicture}
\draw[red,rounded corners]  (0,0)--++(-0.3,0) --++ (0,0.25); 
\draw[red,rounded corners] (2.6,0) --++ (0.3,0)--++(0,0.2); 
\foreach \x in {0,1.3}{
\pic[scale=0.6] (v) at  ($ (\x,0) + (0.5,0) $) {V1p=//};
\draw[red] (v-mid) -- (\x,0); 
    \node[tensor] (t) at (\x,0) {};
    \draw (t)-++ (0,0.3)-++ (0,-2);
\draw[red,rounded corners] (v-up)--++(0.125,0)--++(0,0.25);
\draw[red] (v-down) to [out=0,in=180] ($(\x,0) + (1.3,0)$);
}
\draw (2.6,0.3)--(2.6,-2);
\node[tensor] at (2.6,0) {};
 \node[irrep] at (0.2,0) {$g_1$};
 \node[irrep] at (0.9,-0.1) {$g_2$};
 \node[irrep] at (2.2,-0.1) {$g_3$};
   \node[irrep] at (0.6,0.4) {$g_1\myinv{g}_{2}$};
    \node[irrep] at (1.9,0.4) {$g_{2}\myinv{g}_{3}$};
\draw[rounded corners, fill=white]  (-0.5,-0.35) rectangle (2.9,-0.65);
\node[] at (1.1,-0.5) {$ O $};
\foreach \x in {0,1.3}{
\pic[scale=0.6] (w) at  ($ (\x,-1) + (0.8,0) $) {W1p=//};
\draw[red] (w-mid) -- ($(\x,-1) + (1.3,0)$);
\pic[scale=0.6] (v) at  ($ (\x,-1.6) + (0.5,0) $) {V1p=//};
\node[tensor] (t) at (\x,-1) {};
\draw[red] (w-up) to [out=180,in=0] (t); 
\pic[scale=0.6] (v) at  ($ (\x,-1.6) + (0.5,0) $) {V1p=//};
\node[tensor] (t1) at (\x,-1.6) {};
\draw[red] (w-down) to [out=180,in=0] (v-up);
\draw[red] (v-mid) -- (t1);
\draw[red] (v-down) to [out=0,in=180] ($ (\x+1.3,-1.6) $);
}
\node[tensor] at (2.6,-1) {};
\node[tensor] at (2.6,-1.6) {};
\pic[scale=0.6] (vf) at (3.1,-1.6) {V1p=//};
\pic[scale=0.6] (vi) at (-0.5,-1.6) {V1p=//};
\draw[red] (vf-up) to [out=0,in=0] (2.67,-1);
\draw[red] (vf-mid) -- (2.67,-1.6);
\draw[red] (vi-up) to [out=0,in=180] (-0.07,-1);
\draw[red] (vi-down) to [out=0,in=180] (-0.07,-1.6);
\draw[red] (vi-mid) --++ (-0.6,0);
\draw[red] (vf-down) to [out=0,in=180] (3.5,-1.6);
\draw[red] (-0.9,-1.6) --++ (0.3,0);
\draw (-0.9,-1.4) -- (-0.9,-2);
\draw (3.5,-1.4) -- (3.5,-2);
\draw[red] (3.5,-1.6) --++ (0.2,0);
\node[tensor] at (3.5,-1.6) {};
\node[tensor] at (-0.9,-1.6) {};
\draw[rounded corners, fill=white]  (-1.3,-2) rectangle (3.9,-2.4);
\node[] at (1.1,-2.2) {$ \psi $};
 \node[irrep] at (-0.2,-1.0) {$\myinv{g}_1$};
 \node[irrep] at (1.1,-1.05) {$\myinv{g}_2$};
 \node[irrep] at (2.4,-1.05) {$\myinv{g}_3$};
  \node[irrep] at (0.2,-1.6) {${k}_1$};
  \node[irrep] at (0.9,-1.7) {${k}_{2}$};
    \node[irrep] at (2.2,-1.7) {${k}_{3}$};
\node[irrep] at (-1.2,-1.6) {${k}_{0}$};
\node[irrep] at (3.7,-1.6) {${k}_{4}$};
    \draw[fill=black] (-0.23,0) rectangle (-0.23-0.1,0+0.1);
    \node[] at (-0.25,-0.2) {$Z_{g_1}$};
    \draw[fill=black] (2.83,-1.1) rectangle (2.83+0.1,-1.1+0.1);
    \node[] at (2.9,-0.85) {$Z_{g_3}$};
\end{tikzpicture}\ .
\end{equation}

Let us compute first the isolated term 
\begin{align*}
\begin{tikzpicture}
\draw (-0.7,-0.6) -- (-0.7,0.6);
\draw (0.7,-0.6) -- (0.7,0.6);
\node[tensor] (t1) at (-0.7,0.3) {};
\node[tensor] (t2) at (-0.7,-0.3) {};
\node[tensor] (t3) at (0.7,0.3) {};
\node[tensor] (t4) at (0.7,-0.3) {};
\pic[scale=0.6] (w) at  (0.2,0.3) {W1p= \ \ \  \myinv{g_2}/\myinv{g_1} /};
\pic[scale=0.6] (v) at  (-0.2,-0.3) {V1p=k_1 \ \ / /\ \ k_2};
\draw[red] (v-up) to [out=0,in=180] (w-down);
\draw[red] (v-down) to [out=0,in=180] (t4)--++ (0.3,0);
\draw[red] (v-mid) to [out=180,in=0] (t2)--++ (-0.3,0);
\draw[red] (w-mid) to [out=0,in=180] (t3)--++ (0.3,0);
\draw[red] (w-up) to [out=180,in=0] (t1)--++ (-0.3,0);
\end{tikzpicture}
& =
\begin{tikzpicture}
\pic[scale=0.7] (v1) at (0,0) {V1=/\myinv{g_1}/\ k_1 };
\pic[scale=0.7] (v2) at (0.45,-0.21) {V1u=//\ \ k_2};
\pic[scale=0.7] (w1) at (1.4,-0.2) {W1=/\myinv{g_2}/};
\pic[scale=0.7] (w2) at (1,0.01) {W1D=//};
\draw[red] (v2-up) to [out=0,in=180] (w2-down);
\draw[red] (v1-up) -- (w2-up);
\draw[red] (v2-down) -- (w1-down);
\pic[scale=0.7] (w) at (-0.6,0) {W1=/\myinv{g_1}/ k_1\ \ };
\pic[scale=0.7] (v) at (2,-0.2) {V1=/\ \ \myinv{g_2}/\ \ k_2};
\node[tensor] (t1) at (-0.3,0) {};
\node[tensor] (t2) at (1.7,-0.2) {};
\draw (-0.3,-0.3) -- (-0.3,0.3);
\draw (1.7,-0.5) -- (1.7,0.1);
\end{tikzpicture}
\\ & = \delta_{\myinv{g}_1k_1, \myinv{g_2}k_2}
\begin{tikzpicture}
\pic[scale=0.7] (w) at (-0.8,0) {W1=/\myinv{g_1}/ k_1\ \ };
\pic[scale=0.7] (v) at (0.8,0) {V1=/\ \ \myinv{g_2}/\ \ k_2};
 \node[irrep] at (0,0) {$\myinv{g}_1 k_1 $};
\draw[red] (v-mid) -- (w-mid);
\node[tensor] (t1) at (0.4,0) {};
\node[tensor] (t2) at (-0.4,0) {};
\draw (-0.4,-0.3) -- (-0.4,0.3);
\draw (0.4,-0.3) -- (0.4,0.3);
\end{tikzpicture} \ , 
\end{align*}
which helps us to rewrite Eq.\ \eqref{eq:GOGpsi} as
$$
  \sum_{\substack{g_1,g_2,g_3, \{k_i\} \\  \{ k_0 = \myinv{g}_{1}{k}_{1} = \\   \myinv{g}_{2}{k}_{2}= \myinv{g}_{3}{k}_{3}  = k_4 \} }}
\begin{tikzpicture}
\draw[red,rounded corners]  (0,0)--++(-0.3,0) --++ (0,0.25); 
\draw[red,rounded corners] (2.6,0) --++ (0.3,0)--++(0,0.2); 
\foreach \x in {0,1.3}{
\pic[scale=0.6] (v) at  ($ (\x,0) + (0.5,0) $) {V1p=//};
\draw[red] (v-mid) -- (\x,0); 
    \node[tensor] (t) at (\x,0) {};
    \draw (t)-++ (0,0.3)-++ (0,-1.6);
\draw[red,rounded corners] (v-up)--++(0.125,0)--++(0,0.25);
\draw[red] (v-down) to [out=0,in=180] ($(\x,0) + (1.3,0)$);
}
\draw (2.6,0.3)--(2.6,-1.6);
\node[tensor] at (2.6,0) {};
 \node[irrep] at (0.2,0) {$g_1$};
 \node[irrep] at (0.9,-0.1) {$g_2$};
 \node[irrep] at (2.2,-0.1) {$g_3$};
   \node[irrep] at (0.6,0.4) {$k_1\myinv{k}_{2}$};
    \node[irrep] at (1.9,0.4) {$k_{2}\myinv{k}_{3}$};
\draw[rounded corners, fill=white]  (-0.5,-0.35) rectangle (2.9,-0.65);
\node[] at (1.1,-0.5) {$ O $};
\node[irrep] at (-1.2,-1) {${k}_{0}$};
\node[irrep] at (3.7,-1) {${k}_{4}$};
\draw[red] (-1.2,-1) -- (3.8,-1);
\foreach \x in {-0.9,0,1.3,2.6,3.5}{
\node[tensor] (t) at (\x,-1) {}; }
\draw (-0.9,-0.6) -- (-0.9,-1.6);
\draw (3.5,-0.6) -- (3.5,-1.6);
\draw[rounded corners, fill=white]  (-1.3,-1.4) rectangle (3.9,-1.8);
\node[] at (1.1,-1.6) {$ \psi $};
\draw[fill=black] (-0.23,0) rectangle (-0.23-0.1,0+0.1);
    \node[] at (-0.25,-0.2) {$Z_{g_1}$};
\end{tikzpicture} \ ,
$$
where we have also used Eq.\ \eqref{eq:Zmatrix}. Now we can use the fact that $O$ commutes with the MPO so we obtain
$$
  \sum_{\substack{g_1,g_2,g_3, \{k_i\} \\  \{ k_0 = \myinv{g}_{1}{k}_{1} = \\   \myinv{g}_{2}{k}_{2}= \myinv{g}_{3}{k}_{3}  = k_4 \} }}
\begin{tikzpicture}
\draw[red,rounded corners]  (0,0)--++(-0.3,0) --++ (0,0.25); 
\draw[red,rounded corners] (2.6,0) --++ (0.3,0)--++(0,0.2); 
\foreach \x in {0,1.3}{
\pic[scale=0.6] (v) at  ($ (\x,0) + (0.5,0) $) {V1p=//};
\draw[red] (v-mid) -- (\x,0); 
    \node[tensor] (t) at (\x,0) {};
    \draw (t)-++ (0,0.3)-++ (0,-1.6);
\draw[red,rounded corners] (v-up)--++(0.125,0)--++(0,0.25);
\draw[red] (v-down) to [out=0,in=180] ($(\x,0) + (1.3,0)$);
}
\draw (2.6,0.3)--(2.6,-1.6);
\node[tensor] at (2.6,0) {};
 \node[irrep] at (0.2,0) {$g_1$};
 \node[irrep] at (0.9,-0.1) {$g_2$};
 \node[irrep] at (2.2,-0.1) {$g_3$};
   \node[irrep] at (0.6,0.4) {$k_1\myinv{k}_{2}$};
    \node[irrep] at (1.9,0.4) {$k_{2}\myinv{k}_{3}$};
\draw[rounded corners, fill=white]  (-0.5,-0.9) rectangle (2.9,-1.2);
\node[] at (1.1,-1.05) {$ O $};
\node[irrep] at (-1.2,-1) {${k}_{0}$};
\node[irrep] at (3.7,-1) {${k}_{4}$};
\draw[red,rounded corners] (-1.2,-1) -- (-0.7,-1)-- (-0.7,-0.6)--(3.2,-0.6)--(3.2,-1)--(3.8,-1);

\foreach \x in {0,1.3,2.6}{
\node[tensor] at (\x,-0.6) {}; }
\foreach \x in {-0.9,3.5}{
\node[tensor] at (\x,-1) {}; }
\draw (-0.9,-0.6) -- (-0.9,-1.6);
\draw (3.5,-0.6) -- (3.5,-1.6);
\draw[rounded corners, fill=white]  (-1.3,-1.4) rectangle (3.9,-1.8);
\node[] at (1.1,-1.6) {$ \psi $};
\draw[fill=black] (-0.23,0) rectangle (-0.23-0.1,0+0.1);
    \node[] at (-0.25,-0.2) {$Z_{g_1}$};
\end{tikzpicture}\ .
$$
A direct application of Eq.\ \eqref{eq:calc1}, bearing in mind the identities in the sum's terms, leads us to $\G O | \psi \rangle$, which shows Eq.\ \eqref{compgau}.

Equation \eqref{eq:Uhatgrouprop} is implied by the following:
\begin{align*}
\begin{tikzpicture}
    \pic[scale=1] (v) at (0,0) {V1=gh/g/h};
    \pic[scale=1] (w) at (0.6,-0.6) {W1= \ hn/ /n};
    \pic[scale=1] (v1) at (1.2,-0.6) {V1=//};
    \pic[scale=1] (w1) at (1.8,0) {W1= / /};
     \pic[scale=1] (v2) at (2.4,0) {V1=//};
    \pic[scale=1] (w2) at (3,-0.6) {W1= gh/g /h};
     \node[irrep] at (3.1,0.3) {$gh n \myinv{(gh)}$};
    \draw[red] (v-up)--(w1-up);
    \draw[red] (v1-down)--(w2-down);
\end{tikzpicture} 
 & =
 \begin{tikzpicture}
    \pic[scale=1] (v) at (0,0) {V1=gh/g/h};
    \pic[scale=1] (w) at (0.6,-0.6) {W1= \ hn/ /n};
    \pic[scale=1] (v1) at (1.2,-0.6) {V1=//};
    \pic[scale=1] (w1) at (v1-up) {V1u= / /};
    \pic[scale=1] (w2) at (3,-0.6) {W1= gh/g /h};
    \pic[scale=1] (v2) at (w2-up) {W1u=//};
    \draw[red] (v1-down)--(w2-down);
    \draw[red,rounded corners]  (w1-up)--++ (0.2,0)--++(0,0.3)--(v-up);
    \draw[red,rounded corners]  (v2-up)--++ (-0.2,0)--++(0,0.3)--++(0.7,0);
        \draw[fill=black] (1.8,0.3-0.05) rectangle (1.8+0.1,0.3+0.05);
    \node[] at (1.9,0.45) {$Z_g$};
    \draw[fill=black] (2.4,0.3-0.05) rectangle (2.4+0.1,0.3+0.05);
    \node[] at (2.5,0.45) {$Z_g$};
\end{tikzpicture}
\\ & =
 \begin{tikzpicture}
    \pic[scale=1] (v) at (0,0) {V1=gh/g/h};
    \pic[scale=1] (w) at (0.6,-0.6) {W1= \ hn/ /n};
    \pic[scale=1] (v1) at (1.2,-0.6) {V1u=//};
    \pic[scale=1] (w2) at (2.2,-0.6) {W1u= gh/g /h};
    \draw[red,rounded corners]  (v1-up)--++ (0.2,0)--++(0,0.6)--(v-up);
    \draw[red,rounded corners]  (w2-up)--++ (-0.3,0)--++(0,0.6)--++(0.7,0);
        \draw[fill=black] (1.2,0.3-0.05) rectangle (1.2+0.1,0.3+0.05);
    \node[] at (1.3,0.45) {$Z_g$};
    \draw[fill=black] (2.2,0.3-0.05) rectangle (2.2+0.1,0.3+0.05);
    \node[] at (2.3,0.45) {$Z_g$};
\end{tikzpicture}
\\ & =
 \begin{tikzpicture}
    \pic[scale=1] (w) at (0.6,-0.6) {W1= \ /gh /n};
    \pic[scale=1] (v1) at (1.2,-0.6) {V1=/ \ \ \ \ \ \ \ \ \ gh n \myinv{(gh)}/gh};
\end{tikzpicture}
,
\end{align*}
where we have used Eqs.\ \eqref{eq:Zmatrix} and \eqref{3cocytriv} repeatedly. 

To prove Eq.\ \eqref{eq:UhatGNprop} we use an expression for the gauging map $\G_N$ analogous to Eq.\ \ \eqref{eq:gaugeop} and directly compute the following expression:
\begin{align*}
\hat{U}_g \G_N & = \sum_{\{ n_i\in N\}} \cdots
 \begin{tikzpicture}
 \node[tensor] (t1) at (0,0.05) {};
 \node[tensor] (t2) at (0,-0.6) {};
 \draw (0,0.3)--(0,-0.9);
\draw[red] (t2) --++ (-0.3,0);
    \pic[scale=0.8] (v) at (0.4,-0.6) {V1u=//};
    \pic[scale=0.8] (w) at (0.9,-0.2) {W1p= / /};
    \draw[red] (w-up)-- (t1) --++ (-0.3,0); 
    \pic[scale=0.8] (v1) at (1.3,0) {V1p=//};
    \draw[red,rounded corners] (v1-up)--++ (0.1,0)--++ (0,0.2); 
    \draw[red] (w-mid) to [out=0,in=180] (v1-mid);
     \node[tensor] (t3) at (1.8,0.05) {};
     \draw[red] (v1-down) to [out=0,in=180] (t3)--++ (0.3,0);
 \node[tensor] (t4) at (1.8,-0.6) {};
 \draw[red] (w-down) to [out=180,in=0] (v-up);
 \draw[red] (v-down) to [out=0,in=180] (t4) --++ (0.3,0); 
 \draw (1.8,0.3)--(1.8,-0.9);
 \node[irrep] at (-0.2,0) {$g$};
  \node[irrep] at (2,0) {$g$};
   \node[irrep] at (-0.2,-0.6) {$n_1$};
  \node[irrep] at (2,-0.6) {$n_2$};
\end{tikzpicture}
\cdots
\\ & = \sum_{\{ n_i\in N\}} \cdots
 \begin{tikzpicture}
    \pic[scale=1] (v) at (0.5,0) {V1m=g n_1/g/n_1};
    \pic[scale=0.8] (v1) at (v-down) {V1u=//};
    \pic[scale=0.8] (w1) at (1.5,0.15) {W1p=//};
    \pic[scale=0.8] (v2) at (2,0.3) {V1p=//};
    \pic[scale=1] (w2) at (2.7,0) {W1p=\ g n_2/g/n_2};
    \draw[red] (w1-up) to [out=180,in=0] (v-up);
    \draw[red] (w1-down) to [out=180,in=0] (v1-up);
    \draw[red] (w1-mid) to [out=0,in=180] (v2-mid);
    \draw[red] (w2-up) to [out=180,in=0] (v2-down);
     \draw[red] (w2-down) to [out=180,in=0] (v1-down);
     \draw[red,rounded corners] (v2-up)--++ (0.1,0)--++ (0,0.2);
     \node[tensor] (t1) at (0,0) {};
 \node[tensor] (t2) at (3.2,0) {};
 \draw[red] (w2-mid) -- (t2) --++ (0.3,0);
 \draw[red] (v-mid) -- (t1) --++ (-0.3,0);
 \draw (0,0.3)--(0,-0.3);
 \draw (3.2,0.3)--(3.2,-0.3);
 \end{tikzpicture}
 \cdots
\\ & =  \sum_{\{ n_i\in N\}} \cdots
 \begin{tikzpicture}[baseline=-2mm]
      \node[tensor] (t1) at (0,0) {};
       \draw (0,0.3)--(0,-0.3);
    \node[tensor] (t2) at (1.3,0) {};
    \draw (1.3,0.3)--(1.3,-0.3);
    \pic[scale=0.8] (v) at (0.7,0) {V1p=g n_1 \ \ \ //};
    \draw[red,rounded corners] (v-up)--++ (0.1,0)--++ (0,0.2);
     \draw[red] (v-mid) -- (t1) --++ (-0.3,0);
 \draw[red] (v-down) to [out=0,in=180] (t2) --++ (0.3,0);
 \node[irrep] at (1.6,0) {$g n_2$};
  \node[irrep] at (0.8,0.5) {$g n_1 \myinv{n}_2 \myinv{g}$};
     \end{tikzpicture}
 \cdots
 \\ & =  \sum_{\{ n_i\in N\}} \cdots
 \begin{tikzpicture}
\pic[scale=1] (w) at (0,0) {W1m=\ gn_1/gn_1\myinv{g}/g};
 \node[tensor] (t1) at (-0.7,0.3) {};
 \node[tensor] (t2) at (-0.7,-0.3) {};
 \pic[scale=0.7] (v) at (0.7,0.2) {V1p=//};
 \pic[scale=1] (v1) at (1.4,0) {V1m=gn_2\ \ / \ gn_2\myinv{g}/g};
  \node[tensor] (t3) at (2.1,0.3) {};
 \node[tensor] (t4) at (2.1,-0.3) {};
 \draw (-0.7,0.6)--(-0.7,-0.6);
  \draw (2.1,0.6)--(2.1,-0.6);
\draw[red] (w-up)--(t1) --++ (-0.3,0);
\draw[red] (w-down)--(t2) --++ (-0.3,0);
\draw[red] (v1-up)--(t3) --++ (0.3,0);
\draw[red] (v1-down)--(t4) --++ (0.3,0);
    \draw[red,rounded corners] (v-up)--++ (0.15,0)--++ (0,0.3); 
    \draw[red] (v-down) to [out=0,in=180] (v1-mid);
    \draw[red] (w-mid) to [out=0,in=180] (v-mid);
\end{tikzpicture}
 \cdots
  \\ & =  \sum_{\{ n_i\in N\}} \cdots
 \begin{tikzpicture}
\pic[scale=1] (w) at (0,0) {W1m=\ \  gn_1/gn_1\myinv{g}/g};
 \node[tensor] (t1) at (-0.7,0.3) {};
 \node[tensor] (t2) at (-0.7,-0.3) {};
 \pic[scale=1] (v) at (0.7,0) {V1p=//\ \ \ g};
 \pic[scale=0.7] (v1) at (1.3,0.5) {V1p= / / \ \quad \quad gn_2\myinv{g}};
 \draw[red] (w-mid)--(v-mid);
  \node[tensor] (t3) at (2.1,0.3) {};
 \node[tensor] (t4) at (2.1,-0.3) {};
 \draw (-0.7,0.6)--(-0.7,-0.6);
  \draw (2.1,0.6)--(2.1,-0.6);
\draw[red] (w-up)--(t1) --++ (-0.3,0);
\draw[red] (w-down)--(t2) --++ (-0.3,0);
\draw[red] (v1-down)--(t3) --++ (0.3,0);
\draw[red] (v-down)--(t4) --++ (0.3,0);
    \draw[red,rounded corners] (v1-up)--++ (0.15,0)--++ (0,0.3); 
    \draw[red] (v-up) to [out=0,in=180] (v1-mid);
\end{tikzpicture}
 \cdots
 \\ & =  \sum_{\{ n_i\in N\}} \cdots
 \begin{tikzpicture}
 \node[tensor] (t1) at (-0.4,0.3) {};
 \node[tensor] (t2) at (-0.4,-0.3) {};
 \pic[scale=0.7] (v) at (0.3,0.2) {V1p=//};
  \node[tensor] (t3) at (1,0.3) {};
 \node[tensor] (t4) at (1,-0.3) {};
 \draw (-0.4,0.6)--(-0.4,-0.6);
  \draw (1,0.6)--(1,-0.6);
\draw[red] (v-mid) to [out=180,in=0] (t1) --++ (-0.3,0);
\draw[red] (v-down) to [out=0,in=180] (t3) --++ (0.3,0);
\draw[red] (-0.7,-0.3)-- (t2) -- (t4) --++ (0.3,0);
    \draw[red,rounded corners] (v-up)--++ (0.15,0)--++ (0,0.3); 
    \node[irrep] at (-0.8,0.3) {$g n_1 \myinv{g}$};
    \node[irrep] at (-0.6,-0.3) {$g $};
        \node[irrep] at (1.4,0.3) {$g n_2 \myinv{g}$};
    \node[irrep] at (1.2,-0.3) {$g $};
\end{tikzpicture}
 \cdots \ ,
\end{align*}
which is equal to $\G_N {U}_g$ since   conjugation with $g$ on $N$ is an automorphism and amounts to  relabeling the elements in the sum. We notice that in the last equality we have used the following property
$$ 
\begin{tikzpicture}
    \pic[scale=0.8] (w) at (-0.25,0) {W1=/h/g};
    \pic[scale=0.8] (v) at (0.25,0) {V1=hg \ /h/g};
\end{tikzpicture} =
\begin{tikzpicture}
    \draw[red] (-0.3,0.25) --++ (0.6,0);
    \draw[red] (-0.3,-0.25) --++ (0.6,0);
    \node[] at (0,0.4) {$h$};
    \node[] at (0,-0.13) {$g$};
\end{tikzpicture} \ ,
$$
for $h= gn\myinv{g}$.

\section{MPO fusion category symmetries}\label{sec:gauFCsym}
Recently, there has been increased interest in studying 1D systems with more general anomalous symmetries  \cite{Thorngren19,Aasen20, Garre22}. Those symmetries do not necessarily form a group, rather they are representations of fusion categories \cite{EGNObook}. For example, the Fibonacci fusion category which is given by two objects: the trivial element $1$ and $\tau$ such that $\tau \times \tau = 1 + \tau$ corresponds to the only non-trivial product. Fusion categories can also be represented in the form of MPOs and they arise naturally as the boundary symmetries of PEPS exhibiting intrinsic topological order \cite{Schuch10,Sahinoglu14, Bultinck17A}. One can consider states at the boundaries of such systems whose  symmetries are then MPO representations of  fusion categories.

In this section we consider MPOs that represent fusion categories. We show how we can construct local operators with the same algebra relations as the MPOs and how to use them to symmetrize a state. We then particularize for the case of MPSs symmetric under these MPOs.

\subsection{MPOs from fusion categories}

We start by listing the main properties of fusion categories that we use, for a formal exposition see Ref.\cite{EGNObook}.

A fusion category $\mathcal{C}$ is a set of simple objects labeled by $a,b,c,$ etc with a product, called fusion, that satisfy $a\times b= \sum_c N_{ab}^c c$ where $N_{ab}^c$ is the fusion multiplicity. There is an identity object, labeled by $1$, that fuses trivially with the others and that for every object $a$, there is an $\bar{a}\in \mathcal{C}$ such that $N_{a\bar{a}}^1\neq 0$. The multiplicities satisfy $N_{ab}^c= N_{\bar{a}b}^c =N_{ac}^b$ and we assume for simplicity that $N_{ab}^c = \{ 0,1\}$, the non-multifusion case. To every object $a$, the positive value $d_a=d_{\bar{a}}$ can be assigned, called quantum dimension, and they satisfy $d_a d_b = \sum_c N_{ab}^c d_c$ and $\sum_{a,b} N_{ab}^c d_a d_b = \mathcal{D}^2 d_c$, where $\mathcal{D}^2= {\sum_a d_a^2}$ is the total dimension squared. These properties allow one to define the element $\Lambda=\sum_a \frac{d_a}{\mathcal{D}^2} a $ satisfying $\Lambda^2 = \Lambda$ and  $a\times \Lambda = \Lambda \times a = d_a \Lambda$ that we will use later on.

MPO representations of fusion categories have been studied in Refs.\ \cite{Sahinoglu14,Bultinck17A}. We denote by $O_a$ the MPO representing the object $a \in \mathcal{C}$ and we represent it graphically as
\begin{equation*}
O_a =
			\begin{tikzpicture}
		  \draw[red,rounded corners] (0.25,0.3) rectangle (2.75,-0.2);
		  \foreach \x in {0.75,2.25}{
		    \node[tensor] (t\x) at (\x,0.3) {};
		    \draw (\x,0) --++ (0,0.6);
      }
       \node[] at (0.75+0.25,0.5) {$a$};
		  \node[fill=white] at (1.5,0.3) {$\dots$};
		\end{tikzpicture}   \ .
\end{equation*}
The relation $O_a O_b = \sum_c N_{ab}^c O_c$ is satisfied by the existence of fusion tensors that fulfill the following:

\begin{equation}\label{fusiontenC}
  \begin{tikzpicture}[scale=0.75]
    \draw[red]  (-0.5,0)--(0.5,0);
    \draw[red]  (-0.5,-0.5)--(0.5,-0.5);
    \node[irrep] at (-0.35,0) {$a$};
    \node[irrep] at (-0.35,-0.5) {$b$};
    \node[tensor] (t) at (0,0) {};
    \node[tensor] (t) at (0,-0.5) {};
    \draw (0,-0.8) -- (0,0.3);
  \end{tikzpicture} =
  \sum_{c}
  \begin{tikzpicture}[baseline=-1mm]
    \node[tensor] (t) at (0,0) {};
    \draw (0,-0.3) -- (0,0.3);
    \pic[scale=0.75] (v) at (0.4,0) {V1=c/a/b};
    \pic[scale=0.75] (w) at (-0.4,0) {W1=c/a/b};
  \end{tikzpicture} \ , \ 
   \begin{tikzpicture}[baseline=-1mm]
    \pic[scale=0.75] (v) at (0,0) {V1=c/a/b};
    \pic[scale=0.75] (w) at (0.45,0) {W1=d//};
   \end{tikzpicture} 
= \delta_{c,d}   \id_c \ .
\end{equation}
In the previous equation we have omitted the factors $N_{ab}^c$ in the sum; we will do so by considering only fusion tensors with $N_{ab}^c\neq 0$. By virtue of associativity of the product the following is satisfied
\begin{equation}\label{Assorel}
 \begin{tikzpicture}[baseline=0mm]
    \pic[scale=0.8] (v1) at (-0.6,0) {V1=d/e/c};
    \pic[scale=0.8] (w1) at (0.6,0) {W1=d/e/c};
     \node[] at (0,0) {$\otimes$};
  \end{tikzpicture} 
  = \sum_f
  \begin{tikzpicture}[baseline=-3mm]
    \pic (v1) at (-1.5,0) {V1=d/a/f};
    \pic (v2) at (v1-down) {V1=/b/c};
    \pic[scale=0.7] (w) at (-0.5,0.21) {W1=e//};
    \draw[red] (v1-up) to [out=0,in=180] (w-up);
    \node[] at (0,0) {$\otimes$};
    \pic (w1) at (1.5,0) {W1=d/a/f};
    \pic (w2) at (w1-down) {W1=/b/c};
    \pic[scale=0.7] (v) at (0.5,0.21) {V1=e//};
    \draw[red] (w1-up) to [out=180,in=0] (v-up);
   \end{tikzpicture}
    .
\end{equation}
It is important to mention the existence of the MPO $O_\Lambda = \frac{1}{\mathcal{D}^2}\sum_a d_a O_a$ which is a projector and satisfies $O_a O_\Lambda =  O_\Lambda O_a=   d_a O_\Lambda$.

\subsection{Localizing and symmetrizing MPOs from fusion category}

As in Section \ref{sec:MPOtopo}, the gauge Hilbert space between sites $i$ and $i+1$ is $\mathcal{H}_{[i]_r} \otimes \mathcal{H}_{[i+1]_l}= \mathbb{C}^\chi \otimes \mathbb{C}^\chi$ where  $\chi=\sum_a {\chi_a}$ corresponds to the virtual dimension of the MPO. We define the following local symmetry operators:
\begin{equation}\label{symopC}
\check{O}_a =\sum_{b,c}
  \begin{tikzpicture}
     \pic[scale=0.6] (v) at (-0.6,0) {V1p=/a/b};
      \pic[scale=0.6] (w) at (0.6,0) {W1p=/a/b};
      \node[irrep] at (0.9,-0.2) {${c}$};
    \node[irrep] at (-0.9,-0.2) {${c}$};
    \draw[rounded corners,red] (v-down)--++ (0.15,0)--++(0,-0.2);
    \draw[rounded corners,red] (v-mid)--++(-0.15,0)--++(0,0.4);
  \draw[rounded corners,red] (w-down)--++(-0.15,0)--++(0,-0.2);
    \draw[rounded corners,red] (w-mid)--++ (0.15,0)--++(0,0.4);
\draw[red] (v-up)--(w-up);
    \node[tensor] (t) at (0,0.175) {};
    \draw (t)-++ (0,0.3)-++ (0,-0.3);
  \end{tikzpicture} \ ,
\end{equation}
which, by virtue of Eq.\ \eqref{Assorel}, are a representation of the fusion category $\mathcal{C}$:
$$ \check{O}_a \check{O}_b = \sum_c N_{ab}^c \check{O}_c .$$

The local projectors that we use to symmetrize are constructed using the element $\Lambda$, so that the projector onto the symmetric subspace is
$$\mathcal{P} = \prod_{i} \check{O}_\Lambda^{[i]}, \  \check{O}_\Lambda = \frac{1}{\mathcal{D}^2}\sum_a d_a \check{O}_a.$$ 

The symmetrization of a state $|\psi\rangle$ is given by 
$$\mathcal{G}_{\phi} | \psi \rangle = \mathcal{P} \left(|\psi \rangle \otimes|\phi \rangle \right ),$$
where $|\phi \rangle $ is any state supported on the gauge Hilbert space. The resulting state $\mathcal{G}_{\phi} | \psi \rangle$ is invariant under $\check{O}^{[i]}_a$, on every site $i$, in the following sense: 
\begin{equation}\label{gaugeCstate}
    \check{O}^{[i]}_a  \cdot \mathcal{G}_{\phi} | \psi \rangle = d_a \cdot  \mathcal{G}_{\phi} | \psi \rangle \ .
\end{equation}
The main difference with the group case is that the states are not completely invariant under the operators, they are invariant up to $d_a$.

\subsection{MPS case}

MPSs invariant under MPOs representing fusion category symmetries have been studied in Ref.\cite{Garre22}. The MPS $|\psi \rangle$ invariant under the MPO $\{ O_a\}$ is a superposition of injective MPSs. We label the injective MPSs by $|\psi_{A_\alpha} \rangle$, where $\alpha$ takes values in $x,y,z,$ etc. The action of the MPOs on every injective MPS is given by 
\begin{equation}\label{symVMPS}
O_a |\psi_{A_x} \rangle =\sum_y M_{a,x}^y |\psi_{A_y} \rangle
\ ,
\end{equation}
where $M_{a,x}^y$ is the action multiplicity and we assume that is either $0$ or $1$ for simplicity. Equation \eqref{symVMPS} is satisfied by the existence of a set of action tensors that decompose the action of the MPO tensor onto the MPS tensor and satisfy the following:
\begin{equation}\label{actiontenC}
  \begin{tikzpicture}[scale=0.75]
    \draw[red]  (-0.5,0)--(0.5,0);
    \draw[]  (-0.5,-0.5)--(0.5,-0.5);
    \node[irrep] at (-0.35,0) {$a$};
    \node[irrep] at (-0.35,-0.5) {$x$};
    \node[tensor] (t) at (0,0) {};
    \node[tensor] (t) at (0,-0.5) {};
    \draw (0,-0.5) -- (0,0.3);
  \end{tikzpicture} =
  \sum_{y}
  \begin{tikzpicture}[baseline=-1mm]
    \node[tensor] (t) at (0,0) {};
    \draw (0,0) -- (0,0.3);
    \pic[scale=0.75] (v) at (0.4,0) {V2=y/a/x};
    \pic[scale=0.75] (w) at (-0.4,0) {W2=y/a/x};
  \end{tikzpicture} \ , \ 
   \begin{tikzpicture}[baseline=-1mm]
    \pic[scale=0.75] (v) at (0,0) {V2=y/a/x};
    \pic[scale=0.75] (w) at (0.45,0) {W2=z//};
   \end{tikzpicture} 
= \delta_{z,y}   \id_y \ .
\end{equation}

An MPS invariant under the MPOs can be constructed by $|\psi \rangle = O_\Lambda |\psi_{A_x} \rangle $, for any $x$, and satisfies the invariant equation:
$$  O_a |\psi \rangle = d_a |\psi \rangle.$$

Let us particularize the gauging procedure for the injective MPS $| \psi_{A_x} \rangle$. For simplicity we take as the initial gauge field state $|V\rangle =\bigotimes_{i}|v\rangle_{[i]_r} |v\rangle_{[i+1]_l}$, where $|v\rangle$ is the vector supported on the unit block $e$ of the MPO satisfying
\begin{equation*}
  \begin{tikzpicture}
      \pic (w) at (0,-0.3) {V1={a}/a/{e}};
    \node[tensor,label= east:$| v \rangle$] (t) at (w-down) {};
  \end{tikzpicture}
=
  \begin{tikzpicture}
    \draw[red] (-0.3,0) -- (0.3,0);
   \node[irrep] at (0,0.0) {${a}$};
  \end{tikzpicture}
  =
    \begin{tikzpicture}
      \pic (w) at (0,-0.3) {W1={a}/a/{e}};
    \node[tensor,label= west:$\langle v |$] (t) at (w-down) {};
  \end{tikzpicture}
\ .
\end{equation*}

Then,
\begin{align*}
\mathcal{G}_{V} | \psi_{A_x} \rangle & = \prod_i \sum_{a_i} \cdots 
  \begin{tikzpicture}
  \draw  (-1,-0.2) -- (1,-0.2);
   \draw[rounded corners,red] (-0.85,0.6)--++(0,-0.3)--++(0.7,0)--++(0,0.3);
      \draw[rounded corners,red] (0.15,0.6)--++(0,-0.3)--++(0.7,0)--++(0,0.3);
  \node[tensor]  at (-0.5,0.3) {};
  \node[tensor,label=below:$A_{x}$]  at (-0.5,-0.2) {};
    \draw[] (-0.5,-0.2)--(-0.5,0.6);
    \node[tensor]  at (0.5,0.3) {};
    \node[tensor,label=below:$A_{x}$]  at (0.5,-0.2) {};
    \draw[] (0.5,-0.2)--(0.5,0.6);
     \node[] at (-0.8,0.12) {$a_i$};
      \node[] at (0.15,0.12) {$a_{\scriptstyle i+1}$};
      \end{tikzpicture}  
      \cdots \\
      & =  \prod_i \sum_{a_i,y_i(a_i,x)} \cdots
         \begin{tikzpicture}
  \draw[red,rounded corners] (-0.5,-0.10)--++(0.4,0)--++(0,0.4);
\draw[red,rounded corners] (0.5,-0.10)--++(-0.4,0)--++(0,0.4);
     \pic[scale=0.6] (v) at (-0.45,-0.3) {V3=//};
      \pic[scale=0.6] (w) at (0.45,-0.3) {W3=//};
\draw (v-down)--(w-down);
\node[irrep] at (0,-0.7) {$x$};
\node[] at (0.3,0.4) {$a_{i+1}$};
\node[] at (-0.35,0.4) {$a_{i}$};
\node[tensor] (t) at (1,-0.30) {};
\draw (t) --++ (0,0.3);
\draw (t) --++ (0.30,0)--++(-0.6,0);
\node[tensor,label=below:$A_{y_{i}}$] (t) at (-1,-0.30) {};
\node[] at (1.2,-0.60) {$A_{y_{i+1}}$};
\draw (t) --++ (0,0.3);
\draw (t) --++ (-0.30,0)--++(0.6,0);
\node[] at (1.6,-0.1) {$\cdots$};
\end{tikzpicture} \ ,
\end{align*}
where in the second equality we have used Eq.\ \eqref{actiontenC}.

\bibliography{bibliography.bib}

\end{document}